\documentclass[preprintnumbers,amsmath,amssymb,floatfix,11pt,prd,onecolumn,
superscriptaddress,nofootinbib]{revtex4-1}

\usepackage{graphicx}
\usepackage{dcolumn}
\usepackage{amsmath,amssymb,mathtools}
\usepackage{epstopdf}
\usepackage{verbatim}
\usepackage[colorlinks=true, citecolor=blue, urlcolor = blue, linkcolor= blue, 
bookmarks=true]{hyperref}

\begin{document}

\title{Quasinormal modes, shadow and greybody factors of 5D electrically 
charged Bardeen black holes}

\author{Kimet Jusufi}
\email{kimet.jusufi@unite.edu.mk}
\affiliation{Physics Department, State University of Tetovo, 
Ilinden Street nn, 1200, Tetovo, North Macedonia}
\affiliation{Institute of Physics, Faculty of Natural Sciences and Mathematics, 
Ss. Cyril and Methodius University, Arhimedova 3, 1000 Skopje, North Macedonia}
\author{Muhammed Amir}\email[]{amirctp12@gmail.com}
\affiliation{Astrophysics and Cosmology Research Unit, School of Mathematics, 
Statistics and Computer Science, University of KwaZulu-Natal, 
Private Bag X54001, Durban 4000, South Africa}
\author{Md Sabir Ali} 
\email{sabir.ali@iitrpr.ac.in} 
\affiliation{Indian Institute of Technology Ropar, Punjab-140001, India}
\author{Sunil D. Maharaj}
\email{maharaj@ukzn.ac.za}
\affiliation{Astrophysics and Cosmology Research Unit, School of Mathematics, 
Statistics and Computer Science, University of KwaZulu-Natal, 
Private Bag X54001, Durban 4000, South Africa}

\date{\today}

\begin{abstract}
We study quasinormal modes (QNMs) in 5D electrically charged Bardeen black 
holes spacetime by considering the scalar and electromagnetic field 
perturbations. The black holes spacetime is an exact solution of Einstein 
gravity coupled to nonlinear electrodynamics in five dimensions, which has 
nonsingular behavior. To calculate QNMs, we use the WKB approximation method 
up to sixth order. Due to the presence of electric charge $q_e > 0$, both 
the scalar and electromagnetic field perturbations decay more slowly 
when compared to the Schwarzschild-Tangherlini black holes. We discover that 
the scalar field perturbations oscillate more rapidly when compared to the 
electromagnetic field perturbations. In terms of damping, the scalar 
field perturbations damp more quickly. Graphically we show that the 
transmission (reflection) coefficients decrease (increase) with an increase 
in the magnitude of the electric charge $q_e$. The emission of gravitational 
waves allows spacetime to undergo damped oscillations due to the nonzero value 
of the imaginary part, which is always negative. The imaginary part of the 
QNMs frequencies is continuously decreasing with an increase in the magnitude 
of the electric charge $q_e$ for a given mode ($l,n$). A connection between 
the QNMs frequencies and the black hole shadow, as well as the geometric 
cross-section in the eikonal limit, is also described.
\end{abstract}
 
\maketitle

\section{Introduction}
\label{intro}
General relativity describes many predictions within its regime and has 
become the most successful theory of gravity. One of the predictions includes 
the existence of black holes, as exact solutions to Einstein field equations. 
Well known black hole spacetimes in general relativity include the exact 
vacuum/electrovacuum solutions, for example, the 
Schwarzschild/Reissner-Nordstr{\"o}m metric in the static case and the 
Kerr/Kerr-Newman metric in the stationary axisymmetric case. These 
solutions contain an unavoidable curvature singularity in their 
interiors \cite{Hawking:1969sw}. The Einstein theory of general 
relativity cannot overcome the existence of singularity in a very high 
curvature regime. To overcome such singular nature, one could consider 
regular or nonsingular black holes as possible viable models. Bardeen 
developed the first model for regular black holes. He studied the collapse 
of charged matter with a repulsive de Sitter core inside the black hole 
instead of its singularity \cite{Bardeen:1968}. The regularity of the 
solution means that for a fixed value of the nonlinear parameters, 
the curvature invariants of the spacetime are finite everywhere, including 
at the origin, $r=0$, assuming the limiting curvature condition is satisfied 
\cite{Frolov:2016pav}. After its first inception, research on singularity-free 
models of black holes continued with significant efforts in the last decades. 
Ay{\'o}n-Beato and Garcia obtained the first exact spherically symmetric 
regular black hole solution in which general relativity is coupled to the 
nonlinear electrodynamics field \cite{AyonBeato:2000zs}. In follow 
up papers, there have been plenty of research work motivated by the idea 
of regularity. Many exact solutions for regular black holes have been 
obtained thus far (see \cite{AyonBeato:1998ub,AyonBeato:1999rg,AyonBeato:1999ec,
Dymnikova:2003vt,Mbonye:2005im,Hayward:2005gi} and references therein). 
Rotating regular black holes have been a test bed in the non-Kerr family of 
black holes \cite{Bambi:2011mj,Bambi:2013qj}. They have been used to study 
several exotic properties ranging from thermal phase transition 
\cite{Ali:2019rjn} to shadow properties \cite{Abdujabbarov:2016hnw,Amir:2016cen}, 
particle acceleration and particle collisions 
\cite{Ghosh:2014mea,Amir:2015pja,Ghosh:2015pra}, geodesics completeness 
\cite{Bambi:2016wdn}, etc. Regular black hole models have been extended to 
higher dimensions in modified theories of gravity including Einstein-Gauss-Bonnet 
\cite{Ghosh:2018bxg,Kumar:2018vsm,Nam:2019clw,Hyun:2019gfz}, Lovelock 
\cite{Aros:2019quj}, and others \cite{Ali:2018boy} to study 
their various geometrical and physical properties.

Black hole perturbations result in the emission of gravitational waves which 
are of of electromagnetic or scalar nature, characterized by some complex 
frequencies called the quasinormal modes (QNMs). The real part of the 
frequencies maintains the oscillations of the gravitational waves while 
the imaginary part leads to damped or undamped features in its nature. When 
one detects such waves emanating from the black hole, one could infer 
relevant information about the particular black hole. The perturbations 
are needed to extract information about the nature of the astrophysical 
black holes. Regge and Wheeler \cite{Regge:1957td} performed the first 
research in this direction and then Zerilli \cite{Zerilli:1974ai} studied 
the radial and polar perturbations of the Schwarzschild black holes. They 
added some small changes onto the unperturbed background, restricting the 
conditions that the energy-momentum tensor is not affected with such 
perturbations. Following these works, Vishveshwara 
\cite{Vishveshwara:1970cc,Vishveshwara:1970zz} and then Chandrasekhar in 
his monograph \cite{Chandrasekhar:1975zza} explored the QNMs explicitly. 
The QNMs have been studied for several spacetimes in general relativity 
as well as in alternative theories of gravity 
\cite{Konoplya:2004xx,Natario:2004jd,Nagar:2005ea,Konoplya:2006rv,
Panotopoulos:2020zbj,Ferrari:2000ep,Li:2001ct}. These modes have also been 
discussed in nonsingular black holes \cite{Toshmatov:2015wga,Panotopoulos:2019qjk}. 
A study of the QNMs in the  context of the AdS/CFT correspondence is also 
important. It is expected that there may be an exciting correlation between 
the thermodynamic properties of the loop quantum black holes and the 
quasinormal ringing of the astrophysical black hole candidates 
\cite{Konoplya:2003ii,Konoplya:2011qq}. It may also be valuable to investigate 
QNMs and a possible connection to the collapsing scenario \cite{Purrer:2004nq}. 

A connection between black holes' QNMs and shadows has attracted much attention 
in recent years as the first image of the black hole was being released. The 
Event Horizon Telescope group recently detected the black hole images using 
the shadow properties \cite{Akiyama:2019cqa,Akiyama:2019fyp,Akiyama:2019eap}. 
There exist plenty of work available in the literature on the 
shadow properties (see \cite{Amir:2018szm,Amir:2018pcu,Jusufi:2019nrn,
Jusufi:2020cpn,Haroon:2019new,Vagnozzi:2020quf,Allahyari:2019jqz,
Vagnozzi:2019apd,Bambi:2019tjh,Khodadi:2020jij,Narang:2020bgo} 
and references therein). The detection of gravitational waves 
\cite{Abbott:2016blz} relating compact objects; such as the black holes 
are one of the outstanding discoveries which motivate us to study the 
marriage between the QNMs and the shadow properties of black holes. 
Therefore, it is very fair and motivating to connect these two properties 
as they open a new arena in black hole physics. The QNMs and the greybody 
factors are two important physical phenomena occurring in the curved 
spacetime background. Furthermore, Hawking radiation is a quantum 
mechanical effect experienced in curved spacetimes has paramount importance 
in the study of black holes related phenomena. Particles emitted from the 
black holes face an effective potential barrier which forces the particles 
to go back into the black hole; a phenomenon called back-scattering 
\cite{Rincon:2018ktz}. The greybody factor is a frequency-dependent quantity 
that measures the deviation from the ideal black body radiation and provides 
us valuable information about the horizon structure and related physics 
\cite{Kanti:2002nr}.  

The main motivation of this work is to study QNMs frequencies and their 
connections to the black hole shadow in the 5D electrically charged Bardeen 
spacetime \cite{Amir:2020fpa}. Along with this, we also find a relationship 
between the QNMs frequencies and the greybody factors. Recently, the study 
connecting the real part of the QNMs frequencies and the shadow radius has been 
explored in the static spacetime \cite{Jusufi:2019ltj,Cuadros-Melgar:2020kqn} 
and in the rotating spacetime \cite{Jusufi:2020dhz}. Therefore, the discussion 
of the present work is twofold. On the one hand, we find the QNMs frequencies 
and on utilizing them, we compute the greybody factors of the black holes. On 
the other hand, by using the eikonal approximation, we develop a relationship 
between the QNMs frequencies and the shadow radius of the black holes. The 
electrically charged solutions are quite important when discussing regular 
black hole solutions. The no-go theorem states that there is no Lagrangian 
function having a Maxwell weak field limit, which gives a solution having a 
regular center. To evade the no-go theorem, the Einstein gravity must be 
coupled to nonlinear electrodynamics with a Lagrangian having a proper Maxwell 
weak field limit. Otherwise, it gives a nontrivial symmetric solution with a 
globally regular spacetime with the proviso that the electric charge is zero 
\cite{Bronnikov:2000vy, Bronnikov:2006fu, Bronnikov:2017tnz, Bronnikov:1979ex}. 
In our Lagrangian, we cannot get the Maxwell limit anywhere, but we are allowed 
to have regular black holes which are electrically charged as permitted by 
the no-go theorem. To investigate the stability of the higher dimensional 
black hole solutions which could exist in nature, we need to derive the 
spectra of the gravitational QNMs. The damped and undamped situations are 
related, respectively, to the stable and unstable black holes. The study of 
the QNMs frequencies have been performed by using many numerical techniques 
including the WKB method \cite{schutz1985black}, Frobenius method 
\cite{Konoplya:2011qq}, method of continued fractions \cite{Leaver:1985ax}, 
the Mashhoon method \cite{blome1984quasi,Ferrari:1984zz}, etc. In our work, 
we follow the WKB approximation method to calculate the QNMs frequencies 
initially developed by the Schutz and Will \cite{schutz1985black}. Iyer and 
Will \cite{iyer1987black}, in their original paper calculated the QNMs up 
to third order. The higher order  contribution was due to Konoplya 
\cite{Konoplya:2003ii} allowing us to compute the QNMs frequencies without 
resorting to complicated numerical methods.  

The paper is organized as follows. We give a short description of  
5D electrically charged Bardeen black holes in Sec.~\ref{spacetime}. 
The QNMs of black holes are comprehensively discussed in 
Sec.~\ref{qnms}. The scattering phenomena and greybody factors are 
the subject of Sec.~\ref{greybf}. Section~\ref{QNMs-shad} is devoted 
to developing a connection between the QNMs frequencies and the shadow 
radius, and the absorption cross-section is discussed in \ref{abcrosec}. 
We end the paper by concluding the key results in Sec.~\ref{concl}.

\section{5D electrically charged Bardeen black holes} 
\label{spacetime}
We begin with a brief description of 5D electrically charged Bardeen 
black holes. The corresponding Einstein-Hilbert action coupled to 
nonlinear electrodynamics can be expressed \cite{Amir:2020fpa} as follows
\begin{equation}\label{action}
   S = \frac{1}{16\pi}\int{d^5{x}\sqrt{-g}\left[R-4\mathcal{L}
    \left(\mathcal{F}\right)\right]},
\end{equation}
where $g$ represents determinant of the spacetime metric, $R$ is the 
Ricci scalar, and $\mathcal{L}(\mathcal{F})$ is the Lagrangian of the 
matter field. Note that the Lagrangian $\mathcal{L}(\mathcal{F})$ is 
a nonlinear function of the electromagnetic 
field strength $\mathcal{F} = F_{\mu\nu} F^{\mu\nu}/4$ with 
$F_{\mu \nu} = \partial_{\mu} A_{\nu} -\partial_{\nu} A_{\mu}$. 
By varying the action with respect to $g_{\mu \nu}$ and $A_{\mu}$, 
one could derive the Einstein field equations and the Maxwell 
equations, respectively. These equations can have the following forms
\begin{eqnarray}\label{einstein}
	R_{\mu\nu}-\frac{1}{2}g_{\mu\nu}R &=& 2\left(\mathcal{L}_{\mathcal{F}} 
	F_{\mu\alpha}F_{\nu}^{\alpha}-g_{\mu\nu}\mathcal{L} \right), 
	\notag \\	[2mm]
	\nabla_{\mu}\left(\mathcal{L}_\mathcal{F}F^{\mu\nu}\right) &=& 0,
\end{eqnarray}
where $\mathcal{L}_\mathcal{F} = \partial \mathcal{L}/\partial 
\mathcal{F}$. The nonlinear electrodynamics sources which describe 
5D electrically charged Bardeen black holes can be given 
\cite{Amir:2020fpa} by the following functions
\begin{eqnarray}
	\mathcal{L}(r) = \frac{\mu q_e^3 \left(3q_e^3 -4r^3\right)}
	{\left(r^3 +q_e^3\right)^{10/3}}, \quad
	\mathcal{L}_{\mathcal{F}}(r) =  \frac{\left(r^3 +q_e^3\right)^{10/3}}
	{7 \mu q_e r^9}. \label{sour}
\end{eqnarray}
It is noticeable that these nonlinear electrodynamics sources are 
expressed as a function of $r$. The corresponding gauge potential is 
given \cite{Amir:2020fpa} by
\begin{equation}
	A^{\mu}= -\frac{\mu r^7}{q_e \left(r^3 + q_e^3\right)^{7/3}} 
	\delta^{\mu}_t,\label{gauge}
\end{equation}
and the field strength of the electromagnetic tensor has the form
\cite{Amir:2020fpa} as follows
\begin{equation}
    \mathcal{F} = - \frac{49 \mu^2 q_e^4 r^{12}}
    {2\left(r^3 + q_e^3\right)^{20/3}}.\label{fstr}
\end{equation}
Thus, the spherically symmetric 5D electrically charged Bardeen 
black holes spacetime \cite{Amir:2020fpa} reads simply
\begin{eqnarray}\label{metric}
	ds^2 &=& -f(r)\; dt^2 + f(r)^{-1}\; dr^2 \notag \\
	&&+r^2 \left(d\theta^2 +\sin^2\theta\; d\phi^2 
	+\cos^2 \theta\; d\psi^2\right),
\end{eqnarray}
where the metric function $f(r)$ has the form
\begin{equation}
	f(r) =  1 - \frac{\mu r^2}{(r^3 +q_e^3)^{4/3}}. \label{mf}
\end{equation}
The parameter $q_e$ represents the nonlinear electric charge and $\mu$ is 
the mass of black holes. The spacetime \eqref{metric} is asymptotically 
flat and it does not contain any spacetime singularity, which could be 
confirmed by computing the curvature scalars \cite{Amir:2020fpa}. In 
other words, we can say that the spacetime \eqref{metric} represents 
the regular black holes having two horizons, but there is no curvature 
singularity. Our next task is to compute the quasinormal modes (QNMs) 
frequencies of the electrically charged Bardeen spacetime.

\section{QNMs of 5D electrically charged Bardeen black holes} 
\label{qnms}
In this section, we are going to study the QNMs in 5D electrically 
charged Bardeen black holes spacetime. QNMs are the solutions 
of the Schr{\"o}dinger-like wave equations that satisfy the boundary 
conditions at the event horizon of black hole and far away from it. 
We compute the black holes' QNMs frequencies by considering the scalar 
field perturbations and the electromagnetic field perturbations.

\subsection{Scalar field perturbations}
\label{sca-per}
\begin{figure*}
	\includegraphics[width=7.9cm]{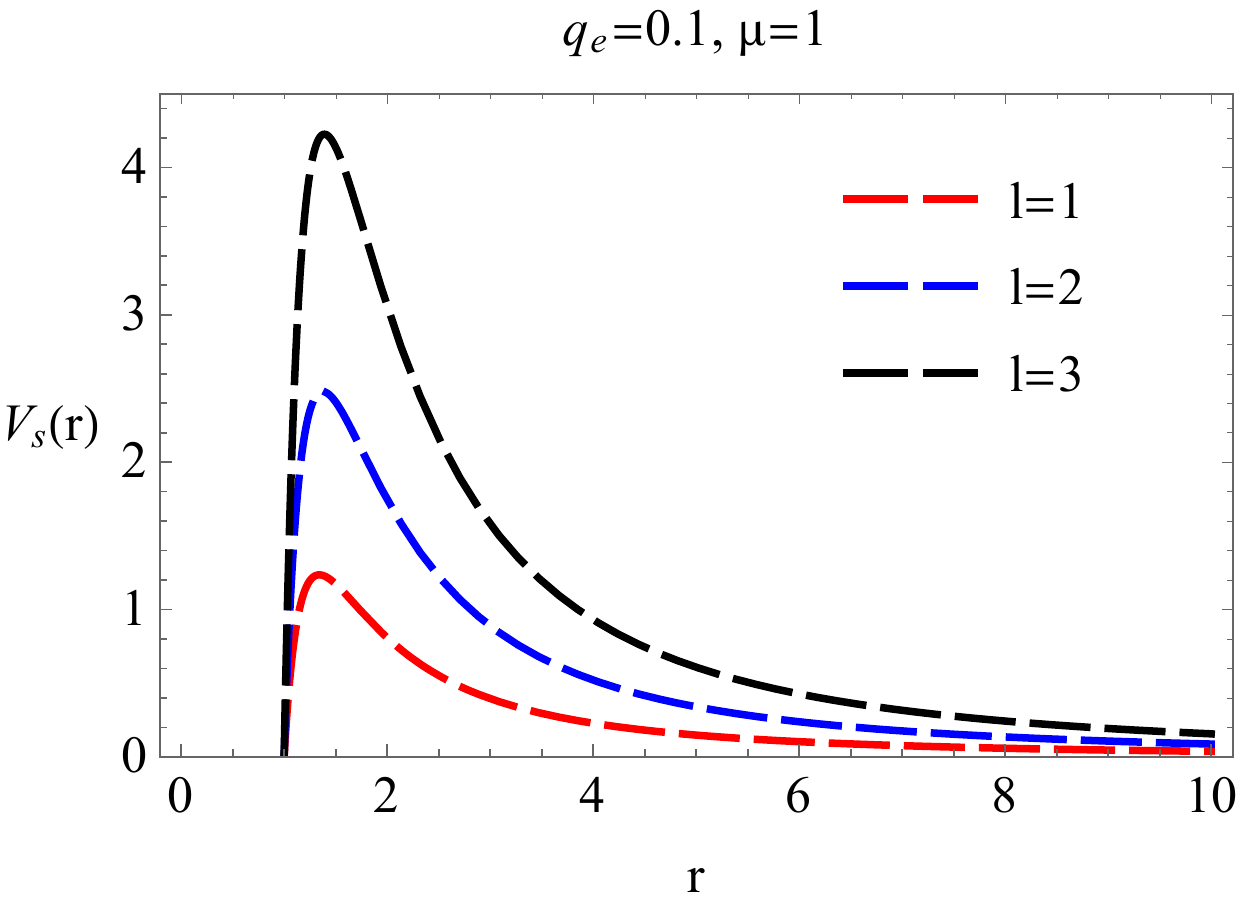}
	\includegraphics[width=7.9cm]{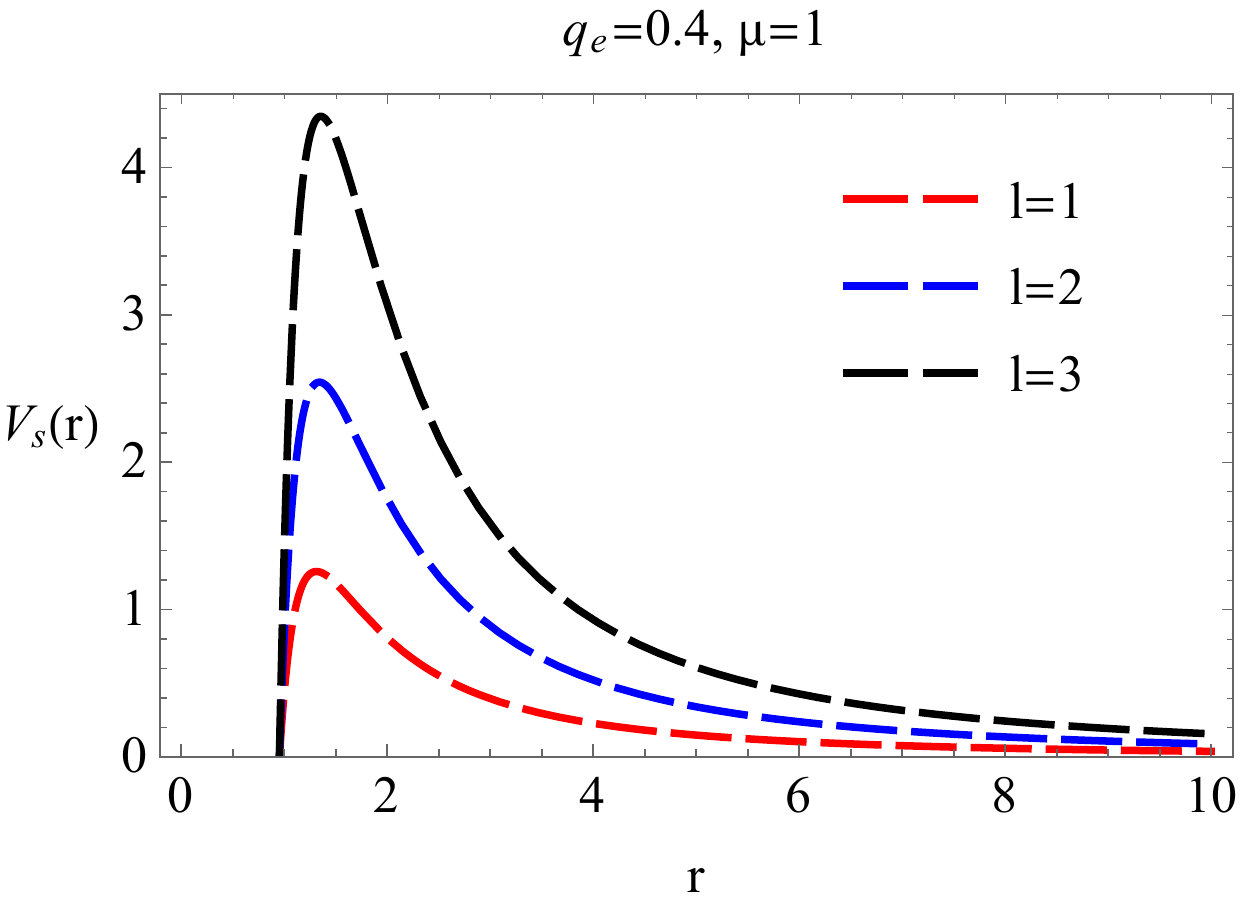}
	\caption{\label{spot} Plot showing the behavior of effective potential 
	function in case of the scalar field perturbations for different 
	values of multipole number $l$ and electric charge $q_e$.}
\end{figure*}

Here we discuss the QNMs of black holes due to the scalar field 
perturbations. We require a solution to the wave equation in order 
to compute the QNMs frequencies. Let us start with equation of motion 
for a massless scalar field which is the Klein-Gordon equation 
\cite{Landau1987}, and in background of the curved spacetime can 
be written as follows
\begin{equation}
	\frac{1}{\sqrt{-g}} \partial_{\mu} \left(\sqrt{-g}\; g^{\mu \nu}\; 
	\partial_{\nu} \Phi \right) =0. \label{kgeq}
\end{equation}
Here $\Phi$ represents the massless scalar field and it is a function 
of coordinates $(t,r,\theta,\phi,\psi)$. We further consider an ansatz 
of the scalar field 
\begin{equation}
	\Phi(t,r,\theta,\phi,\psi) = \sum_{lm} e^{-i\omega t}\; 
	\frac{\Psi_l(r)}{r^{3/2}}\; Y_{lm}(r,\theta) , 
	\label{scfield}
\end{equation}
where $e^{-i\omega t} $ represents the time evolution of field and 
$Y_{lm}(r,\theta)$ denotes the spherical harmonics function. Plunging 
the ansatz \eqref{scfield} into \eqref{kgeq} and applying the separation 
of variables method, we obtain the standard Schr{\"o}dinger-like wave 
equations
\begin{equation}
	\frac{d^2\Psi_l (r_{\ast})}{dr_{\ast}^2} 
	+\left(\omega^2 -V_s(r_{\ast})\right) \Psi_l(r_{\ast})=0, 
	\label{weq}
\end{equation}
where $\omega$ is the frequency of perturbation and $r_{\ast}$ 
represents the tortoise coordinates having the relation
\begin{equation}
	dr_{\ast}=\frac{dr}{f(r)} \Rightarrow r_{\ast} = \int \frac{dr}{f(r)}.
\end{equation}
The advantage of using the tortoise coordinates here is to extend the 
range used in a survey of the QNMs. The tortoise coordinate is being 
mapped the semi-infinite region from the horizon to infinity into 
($-\infty,+\infty$) region. The effective potential in \eqref{weq} 
takes \cite{Molina:2003ff} the following form
\begin{eqnarray}
	V_s(r_{\ast}) &=& \left(1 - \frac{\mu r^2}{(r^3 +q_e^3)^{4/3}}\right)
	\Bigg[\frac{l(l+2)}{r^2} -\frac{6 \mu q_e^3 r^2}{(r^3 +q_e^3)^{7/3}} 
	\notag \\
	&& + \frac{3}{4r^2} \left(1 - \frac{\mu r^2}{(r^3 +q_e^3)^{4/3}}\right)
	\Bigg], \label{seff}
\end{eqnarray}
where $l$ denotes the multipole number. It is noticeable that the 
effective potential \eqref{seff} has the form of a potential barrier.
The typical behavior of the effective potential \eqref{seff} in case 
of the scalar field perturbations can be seen in Fig.~\ref{spot} for 
different values of electric charge $q_e$ and multipole number $l$.

\begin{figure*}
	\begin{center}
		\includegraphics[width=7.9cm]{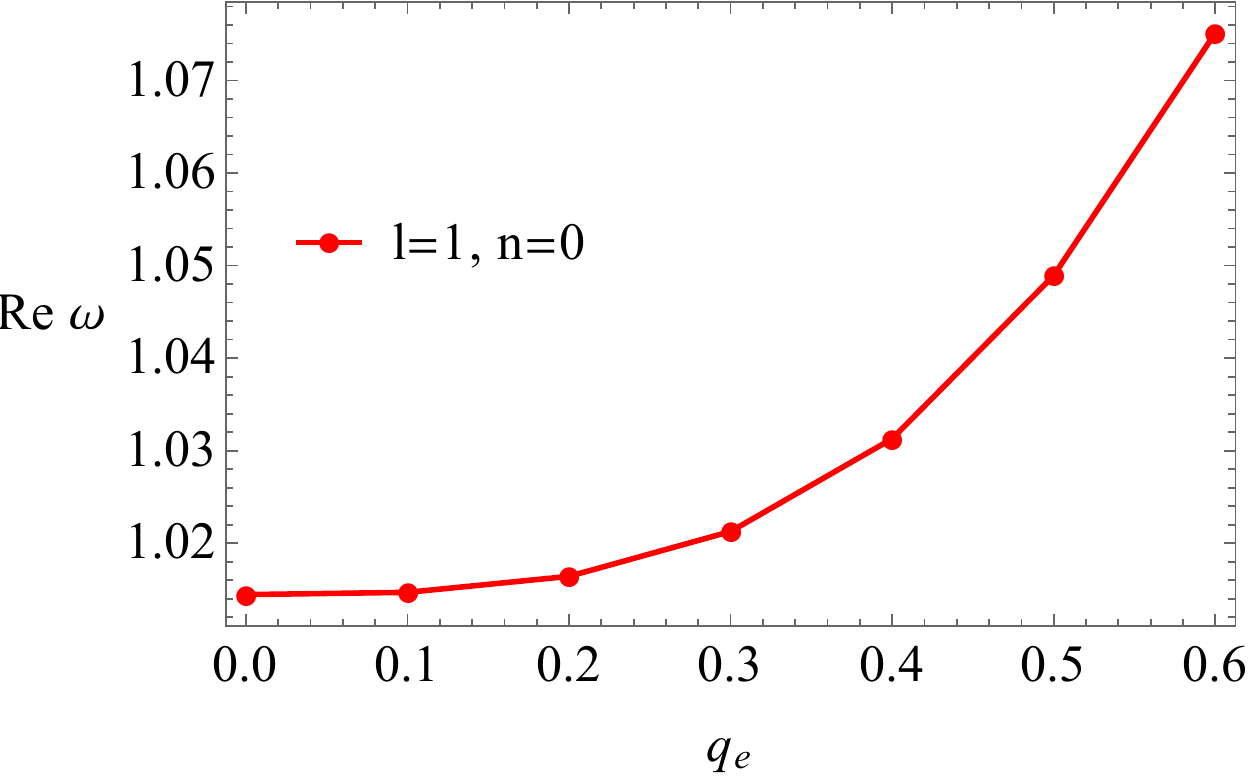}
		\includegraphics[width=7.9cm]{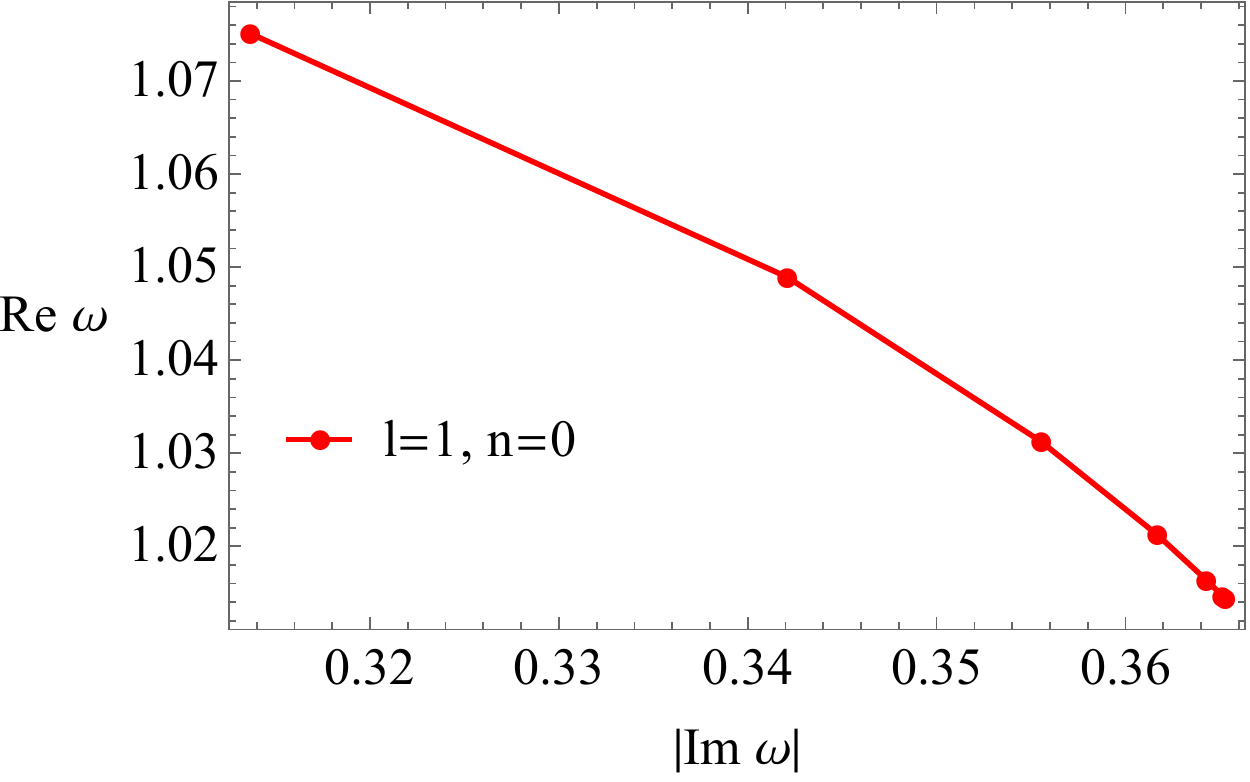}
		\includegraphics[width=7.9cm]{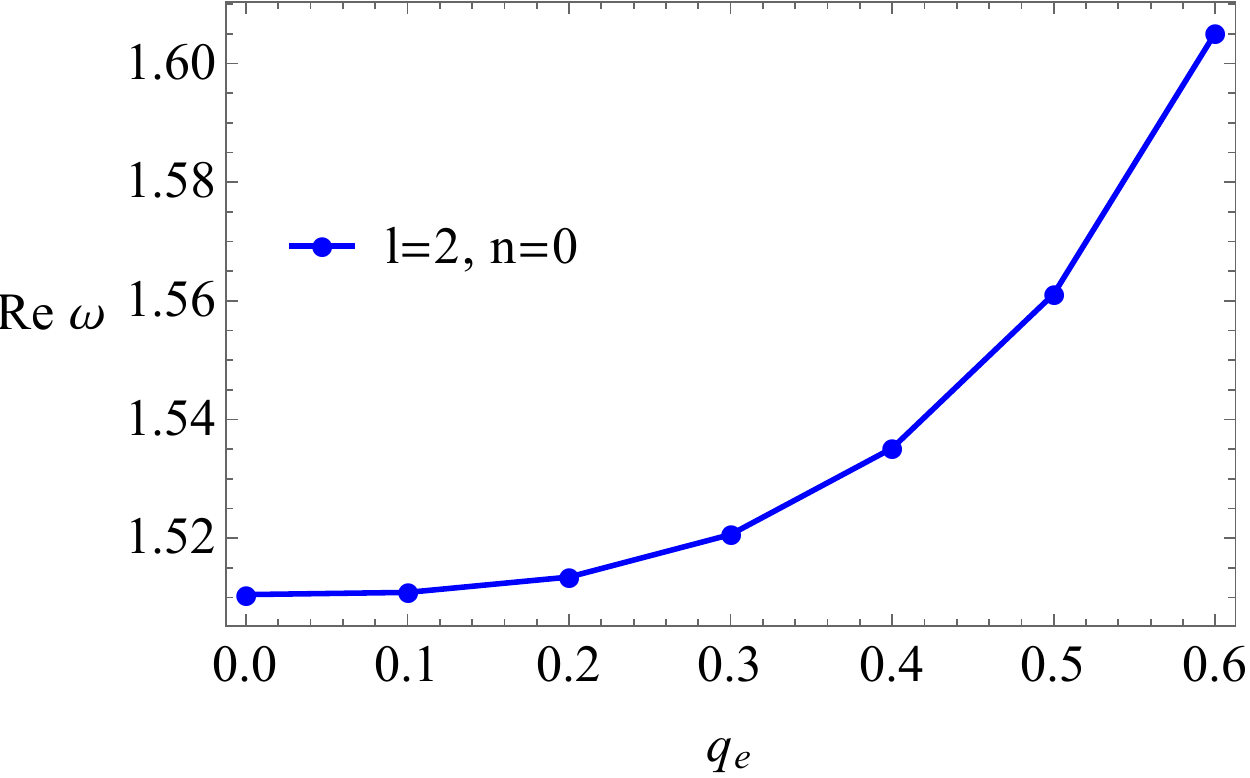}
		\includegraphics[width=7.9cm]{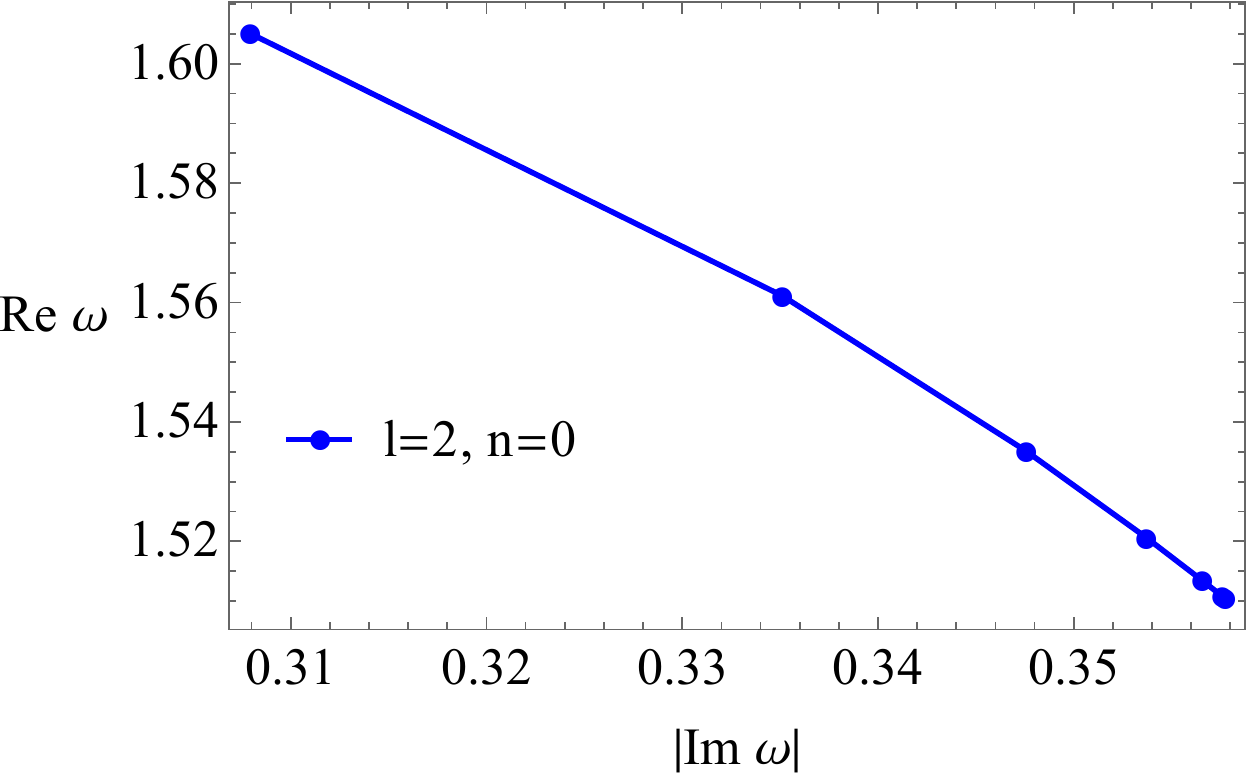}
		\includegraphics[width=7.9cm]{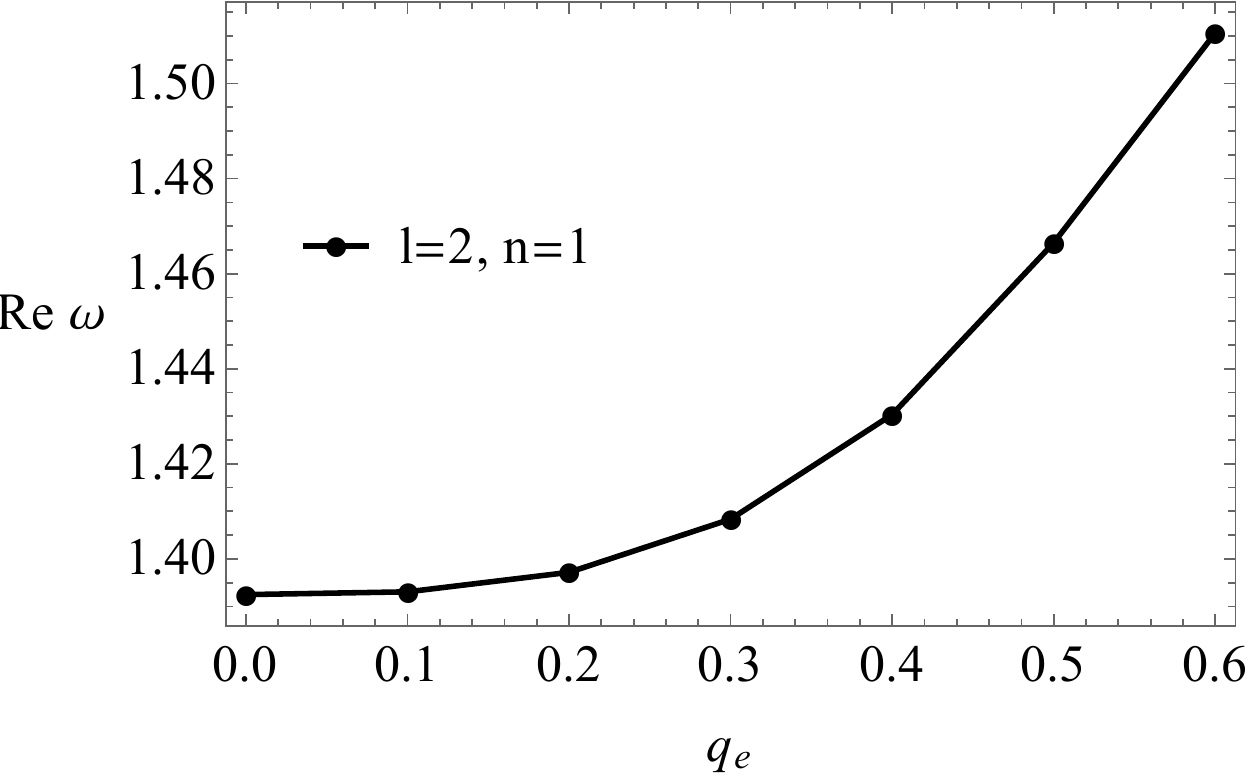}
		\includegraphics[width=7.9cm]{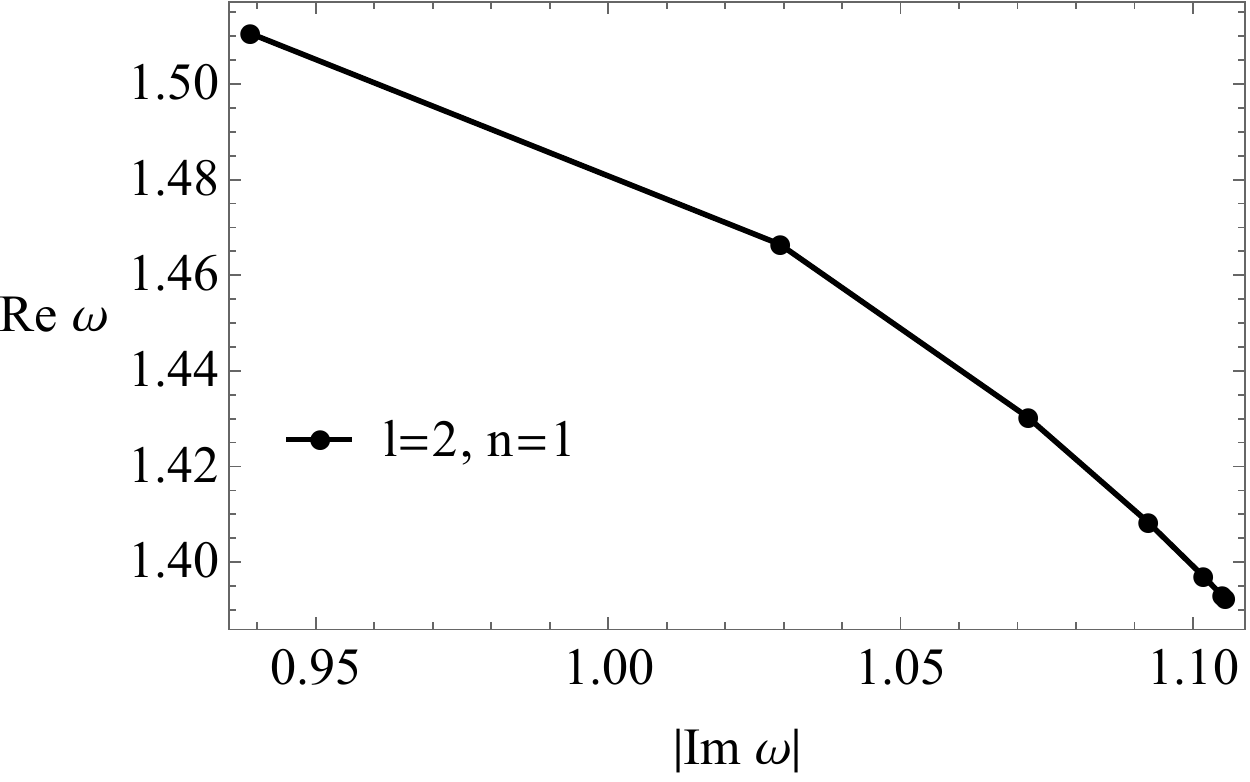}
	\caption{\label{sqnms1} (Left panel) Plots showing the dependence of 
	real part of the QNMs with electric charge $q_e$ for the scalar field 
	perturbations. (Right panel) Plots showing the dependence of real part 
	of the QNMs versus the imaginary part of the QNMs in absolute value
	($\mu=1$).}
	\end{center}
\end{figure*}

\begin{figure*}[ht]
	\includegraphics[width=7.9cm]{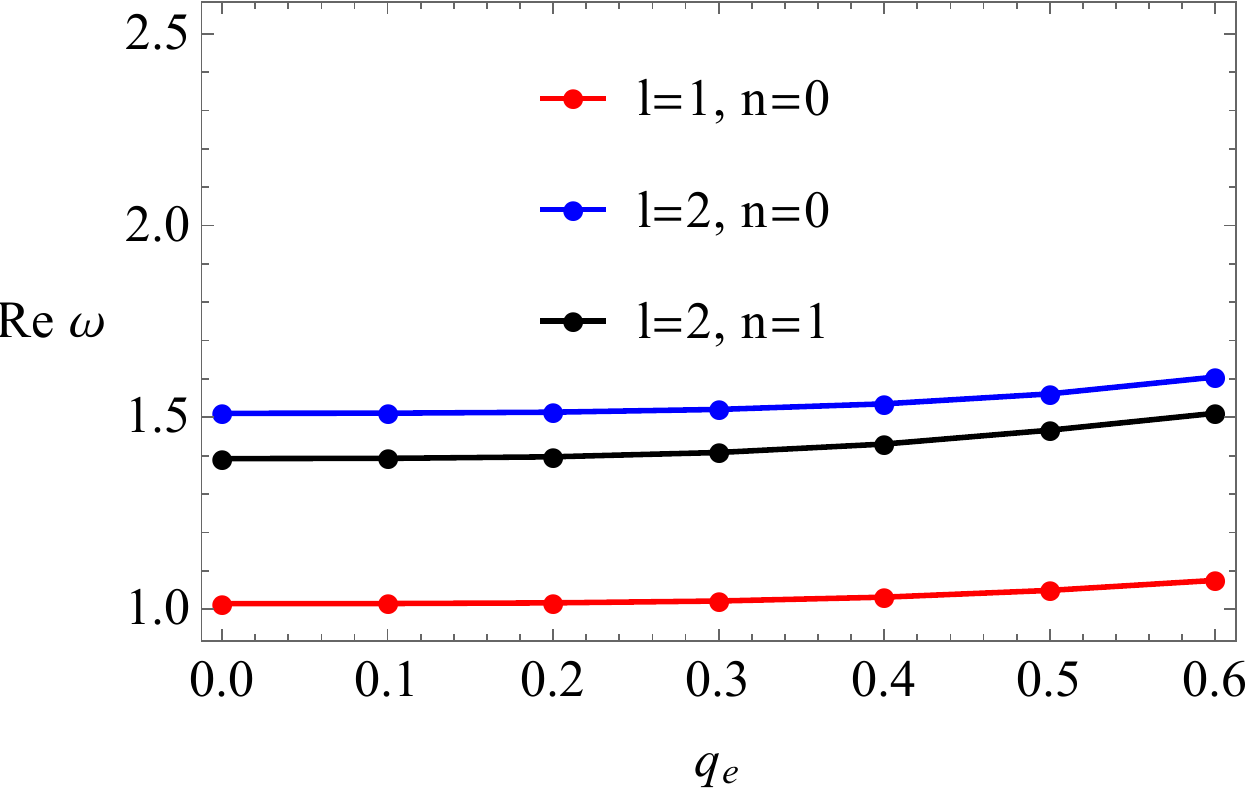}
	\includegraphics[width=7.9cm]{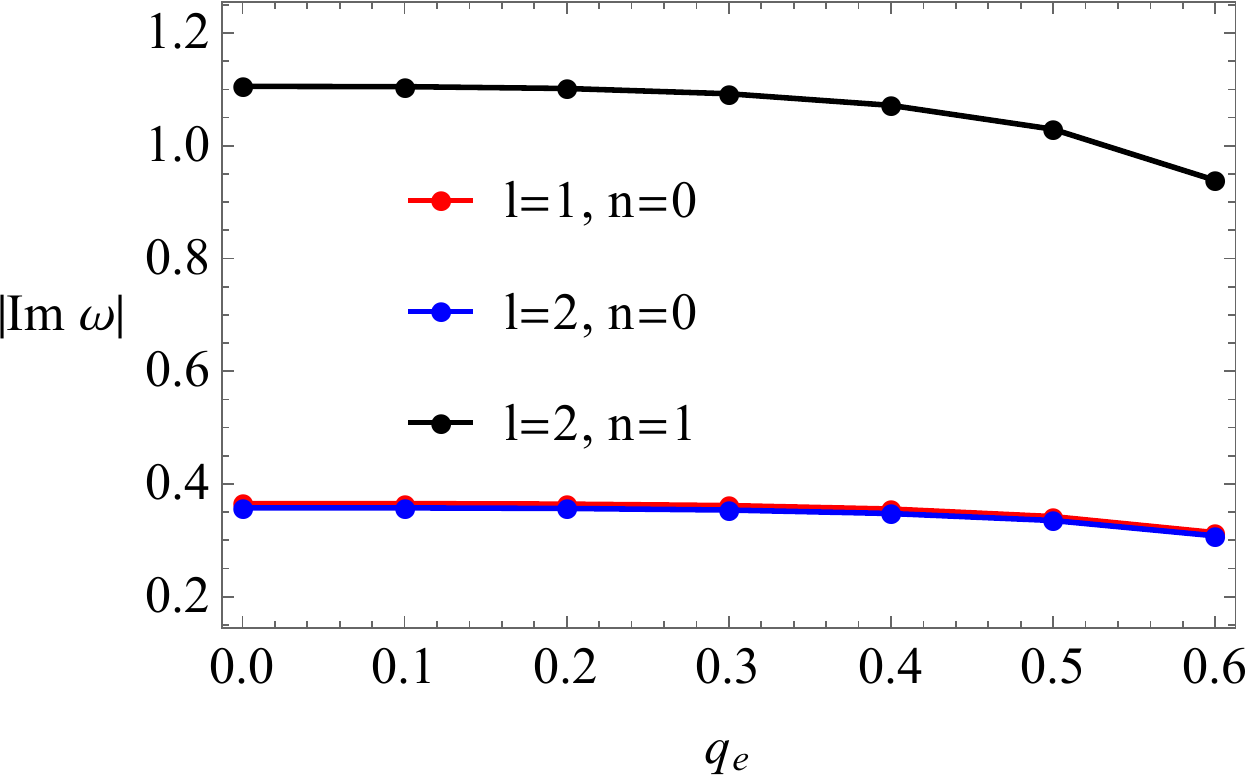}
	\caption{\label{sqnms2} (Left panel) Plot showing the dependence of 
	real part of the QNMs with electric charge $q_e$ for different values 
	of $l$ and $n$ for the scalar field perturbations. 
	(Right panel) Plot showing the dependence of imaginary part of the QNMs 
	in absolute value with electric charge $q_e$ for different values of 
	$l$ and $n$ ($\mu=1$). }
\end{figure*}

On having the expression of the effective potential in our hand, 
we are now in a position to apply the WKB approach in order to compute 
the QNMs due to the scalar field perturbations. The third order QNMs 
frequencies are given \cite{iyer1987black} by
\begin{equation}
	\omega^2=\left(V_0+\sqrt{-2\,V_0''}\; \Lambda_2\right)
	-i \left(n+\frac{1}{2}\right)\sqrt{-2\,V_0''}\; (1+\Lambda_3),
\end{equation}
where $\Lambda_2$ and $\Lambda_3$ are defined as follows
\begin{eqnarray}
	\Lambda_2 &=& \frac{1}{\sqrt{-2\,V_0''}}\Bigg[\frac{1}{8}
	\left(\frac{V_0^{(4)}}{V_0''}\right) 
	\left(\frac{1}{4}+\alpha^2 \right) 
	-\frac{1}{288} \left(\frac{V_0^{(3)}}{V_0''}\right)^2
	\left(7+60\alpha^2 \right)\Bigg], \notag\\
	\Lambda_3 &=& \frac{1}{\sqrt{-2\,V_0''}}\Bigg[\frac{5}{6912}
	\left(\frac{V_0^{(3)}}{V_0''}\right)^4 \left(77+188\alpha^2 \right)
	\notag \\
	&&-\frac{1}{384} \left(\frac{V_0'''^2V_0^{(4)}}{V_0''^3}\right)
	\left(51 +100\alpha^2 \right) 
	+ \frac{1}{2304}\left(\frac{V_0^{(4)}}{V_0''} \right)^2
	\left(67 +68\alpha^2 \right) \notag\\
	&&-\frac{1}{288} \left(\frac{V_0'''V_0^{(5)}}{V_0''^2} \right)
	\left(19 +28\alpha^2 \right)
	-\frac{1}{288}\left(\frac{V_0^{(6)}}{V_0''}\right) 
	\left(5 +4\alpha^2 \right)\Bigg]. \label{lamds}
\end{eqnarray}
On the other hand, $\alpha$ and $V_0^{(m)}$ have the following 
definitions
\begin{eqnarray}
	\alpha=n+\frac{1}{2},\quad V_0^{(m)} 
	=\frac{d^mV}{dr_*^m} \Bigg|_{r_{\star}},
\end{eqnarray}
where $n$ represents the overtone number. In our study, we are going 
to consider the sixth order WKB method which is described in 
\cite{Konoplya:2003ii}. The corresponding formula has the form as 
follows
\begin{equation}
	i\frac{\omega_n^2-V_0}{\sqrt{-2\,V_0''}}
	-\sum_{i=2}^{6}\Lambda_i=n+\frac{1}{2}, \label{sixord}
\end{equation}
where the definitions of higher order corrections 
$\Lambda_4,\;\Lambda_5,\;\Lambda_6$ can be found in \cite{Konoplya:2003ii}. 
Here $V_0$ represents the height of the barrier and $V_0''$ denotes the 
second derivative with respect to the tortoise coordinate of the potential 
at maximum. Note that all the higher order corrections depend on value of 
the potential barrier and its higher derivatives at the maximum. Furthermore, 
we portrait the QNMs frequencies against the electric charge $q_e$, which can 
be seen in Figs.~\ref{sqnms1} and \ref{sqnms2}. We show various cases of the 
QNMs frequencies by considering the different values of multipole number $l$ 
and overtone number $n$ (cf. Figs.~\ref{sqnms1} and \ref{sqnms2}).
\begin{table}[h]
	\caption{\label{table1} Real and imaginary parts of the QNMs frequencies 
	in scalar field perturbations ($\mu=1$).}
	\centering
	\begin{tabular}{|l|l|l|l|l|}
		\hline
	\multicolumn{1}{|c|}{  } &  \multicolumn{1}{c|}{  $l=1, n=0$ } & 
	\multicolumn{1}{c|}{ $l=2, n=0$ } & \multicolumn{1}{c|}{ $l=2, n=1$} \\
		\hline
	$q_e$ & $\omega \,(WKB)$ & $\omega \,(WKB)$ & $\omega \,(WKB)$   \\ 
	\hline
	0 	& 1.01444 - 0.36524 i & 1.51050 - 0.35770 i & 1.39249 - 1.10537 i  \\
	0.1 & 1.01469 - 0.36512 i & 1.51087 - 0.35756 i & 1.39307 - 1.10491 i \\
	0.2 & 1.01641 - 0.36426 i & 1.51346 - 0.35656 i & 1.39717 - 1.10165 i \\
	0.3 & 1.02126 - 0.36167 i & 1.52062 - 0.35371 i & 1.40836 - 1.09229 i \\
    0.4 & 1.03129 - 0.35551 i & 1.53516 - 0.34755 i & 1.43036 - 1.07169 i \\ 
    0.5 & 1.04892 - 0.34209 i & 1.56114 - 0.33511 i & 1.46649 - 1.02934 i \\
    \hline
	\end{tabular}
\end{table}
Also, we present the numerical results of the QNMs frequencies in 
Table~\ref{table1}. We see an increase in the real part of the QNMs 
frequencies with increasing magnitude of the electric charge $q_e$ 
(cf. Table~\ref{table1}). On the other hand, we find that an increase 
in the electric charge $q_e$ decreases the imaginary part of the QNMs 
frequencies in absolute value (cf. Table~\ref{table1}). This indicates 
that the scalar field perturbations in presence of the electric charge 
$q_e > 0$ decay more slowly compared to the Schwarzschild-Tangherlini 
black holes. It is worth noting that we have not calculated the QNMs 
frequencies in case of fundamental mode $l=n=0$ in Table~\ref{table1}. 
This is related to the fact that the WKB method is applicable when 
$l>n$ and does not give a satisfactory degree of precision for this 
fundamental mode ($l=n=0$). Nevertheless, one can apply other methods 
for example the Frobenius method to include this fundamental mode 
(see for example \cite{Konoplya:2018qov}).

\subsection{Electromagnetic field perturbations}
\label{ele-per}
\begin{figure*}
	\includegraphics[width=7.9cm]{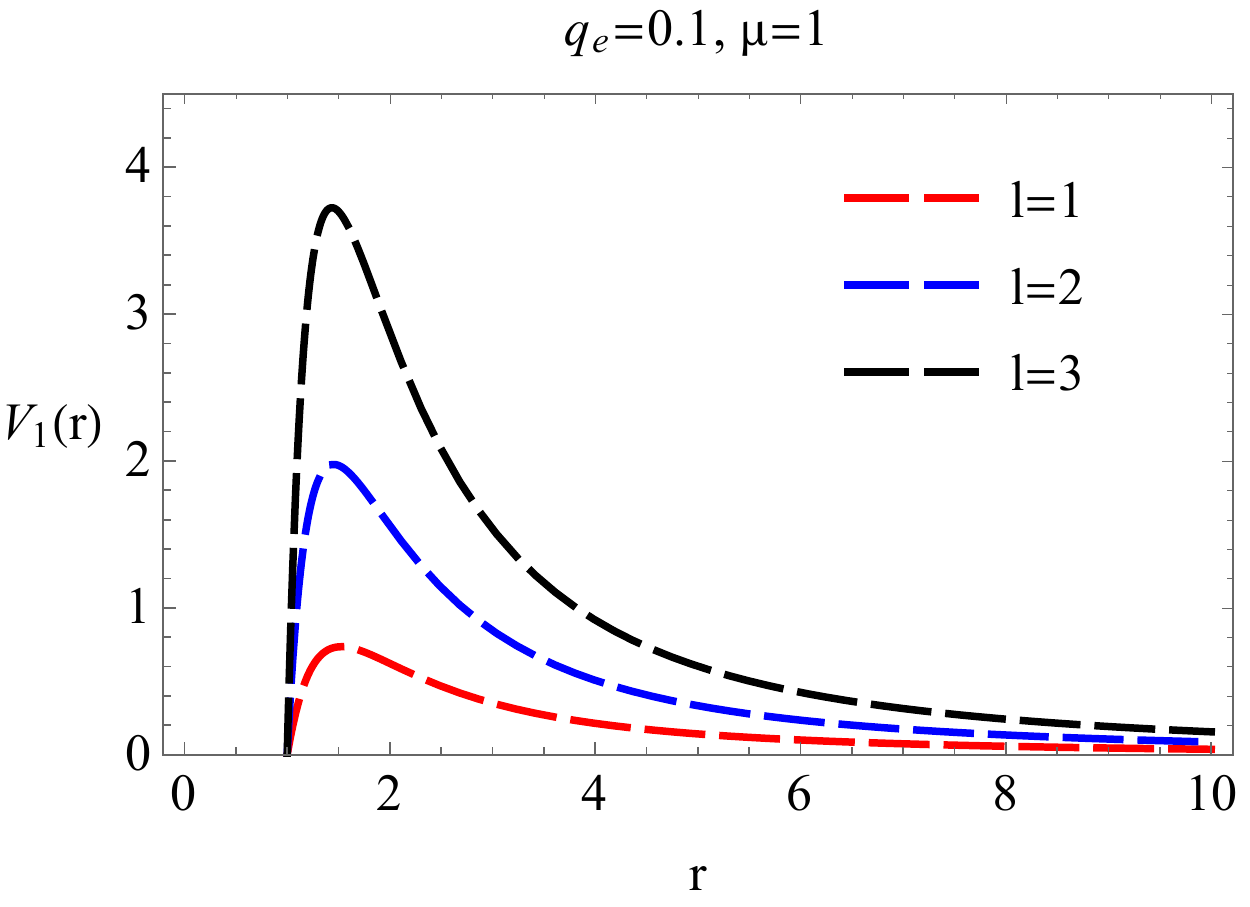}
	\includegraphics[width=7.9cm]{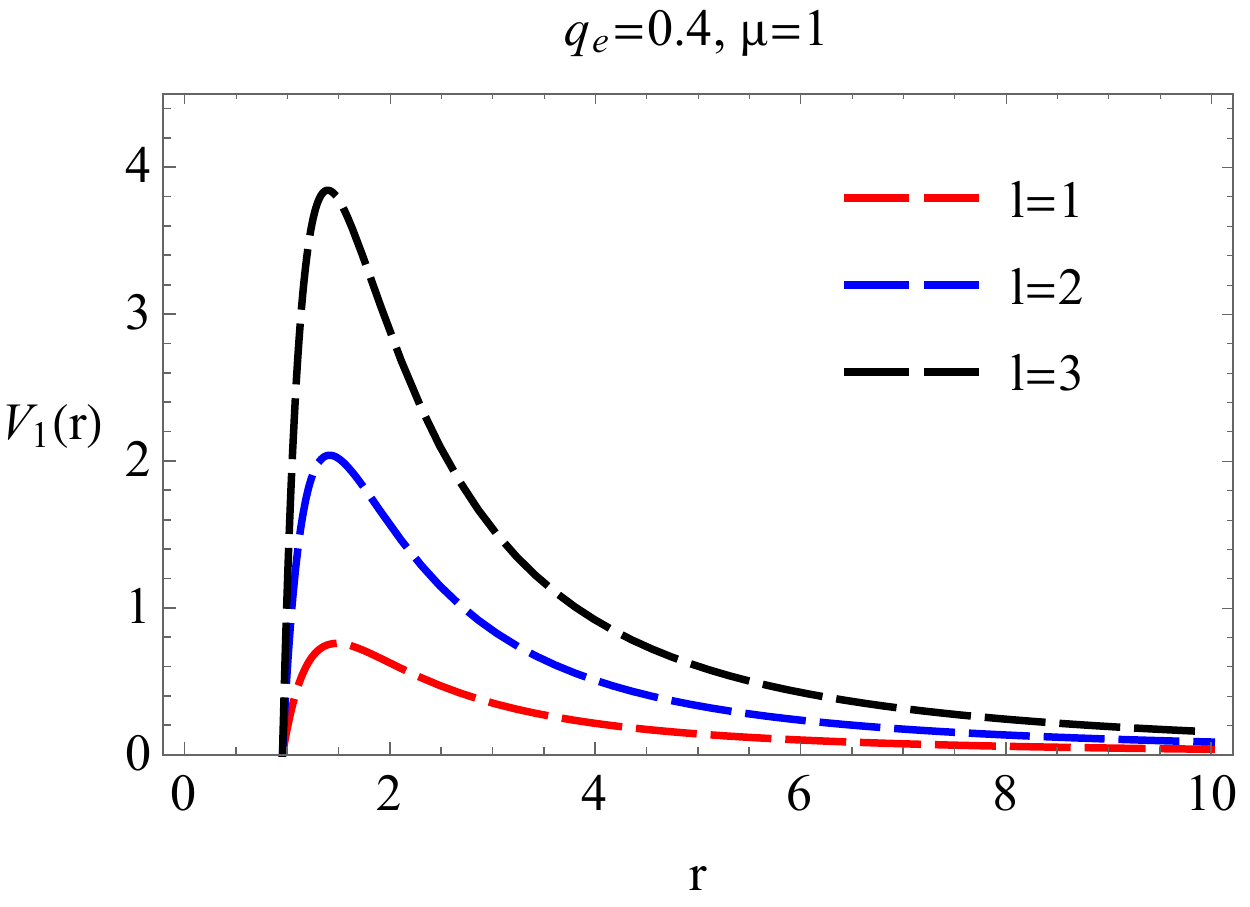}
	\caption{\label{semp} Plot showing the behavior of effective potential 
	function in the case of modes \textbf{I} of the electromagnetic 
	perturbations for different values of multipole number $l$ and electric 
	charge $q_e$.}
\end{figure*}

\begin{figure*}
	\includegraphics[width=7.9cm]{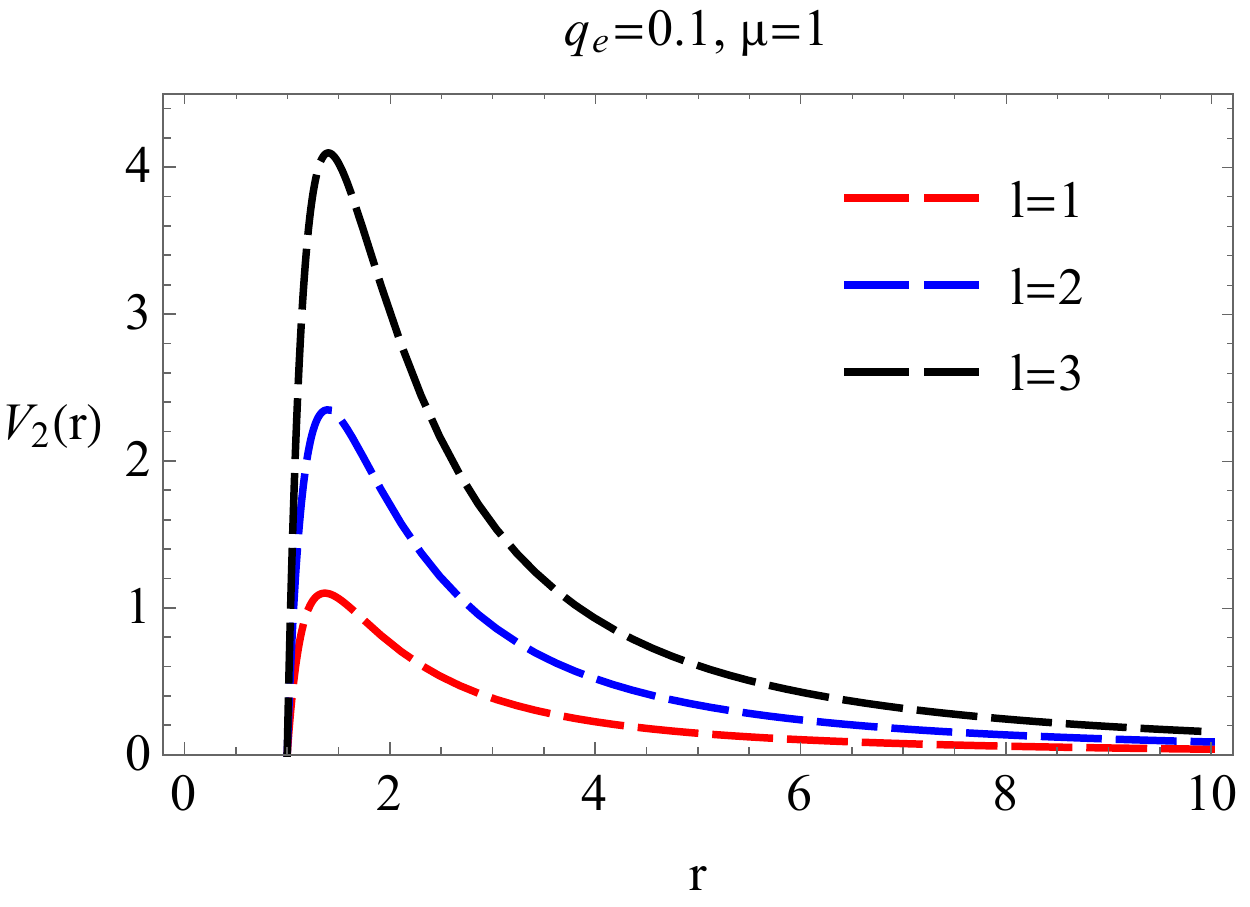}
	\includegraphics[width=7.9cm]{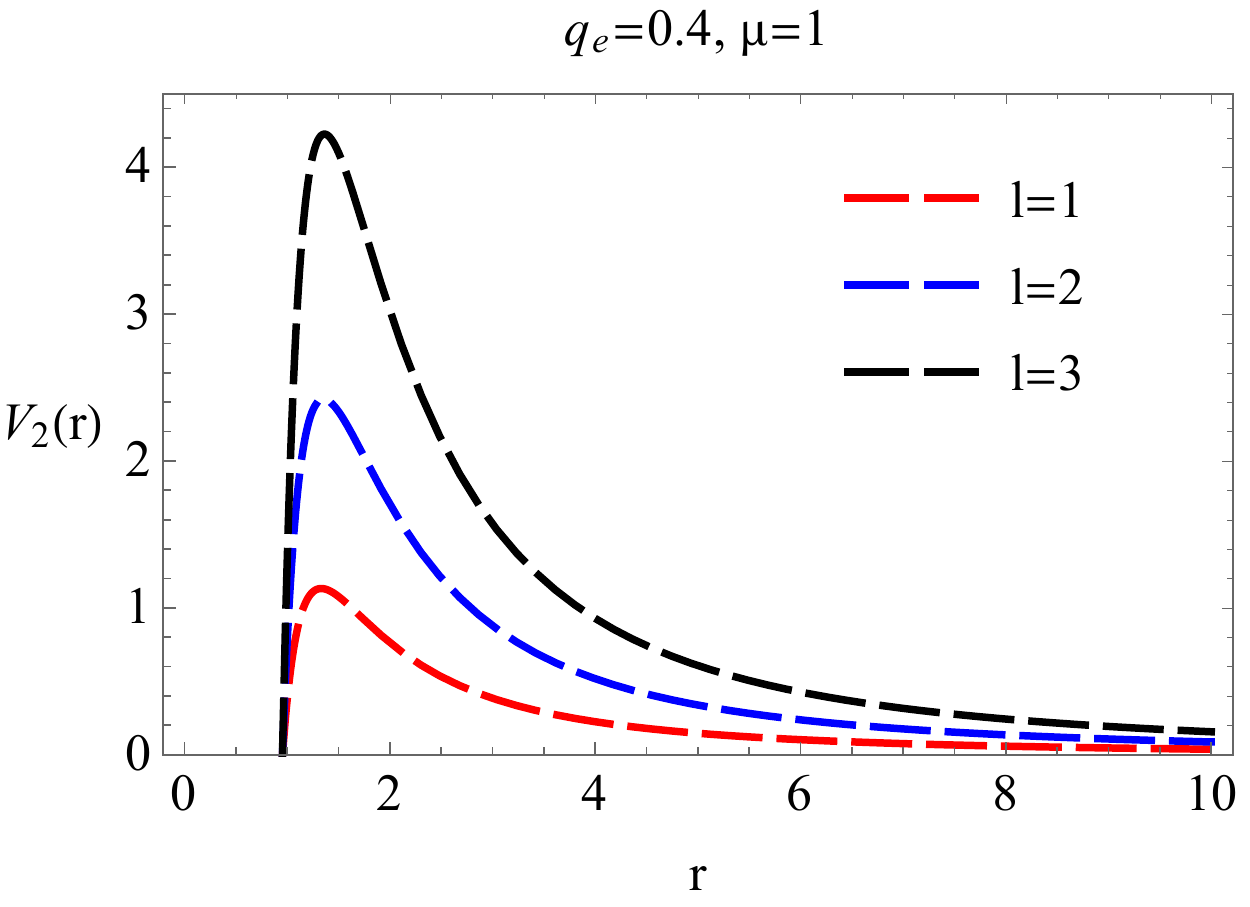}
	\caption{\label{vemp} Plot showing the behavior of effective 
	potential function in the case of modes \textbf{II} of the 
	electromagnetic perturbations for different values of 
	multipole number $l$ and electric charge $q_e$.}
\end{figure*}

In this subsection, we study the QNMs of the black holes by considering 
the electromagnetic field perturbations. To do so we begin with 
rewriting the Maxwell equations, $\nabla_{\mu} 
(\mathcal{L}_{\mathcal{F}} F^{\mu \nu}) =0$ as follows
\begin{equation}
	\frac{1}{\sqrt{-g}}\partial_{\mu}\left[ \sqrt{-g}\; 
	\mathcal{L}_{\mathcal{F}}\;	g^{\lambda \mu}\; g^{\sigma \nu} 
	\left( \partial_{\lambda} A_{\sigma} -\partial_{\sigma} A_{\lambda} 
	\right)	\right]=0. \label{maxeqs}
\end{equation}
Now we substitute all related expressions into \eqref{maxeqs} and using 
the standard tortoise coordinate transformation $dr_{\ast} =dr/f(r)$, which 
turns out to be the second-order differential equations of following form 
\cite{Crispino:2000jx,LopezOrtega:2006vn}:
\begin{equation}
	\frac{d^{2}\Psi ( r_{\ast } ) }{dr_{\ast }^{2}} 
	+\left[ \omega^{2}-V_{1,2} ( r_{\ast }) \right] 
	\Psi ( r_{\ast }) =0.
\end{equation}
Interestingly, we have two categories of the effective potential in case 
of the higher-dimensional black holes while discussing the electromagnetic 
field perturbations {\footnote{Alternatively, one can use the notation 
of Ref. \cite{Kodama:2003kk}, according to which the physical modes 
\textbf{I} and \textbf{II} are noted as scalar type and vector type
electromagnetic perturbations, respectively.}}: i) 
the physical modes \textbf{I} of the electromagnetic field perturbations 
and ii) the physical modes \textbf{II} of the electromagnetic field 
perturbations. Thus, the effective potential corresponding to the 
modes \textbf{I} of the electromagnetic perturbations 
is given \cite{Crispino:2000jx,LopezOrtega:2006vn} by
\begin{eqnarray}
	V_1(r) &=& \left(1 -\frac{\mu r^2}{(r^3 +q_e^3)^{4/3}}\right)\,
	\Big[\frac{l(l+2)}{r^2} \notag\\
	&& + \frac{3}{4r^2}\; 
	\left(1 - \frac{\mu r^2}{(r^3 +q_e^3)^{4/3}}\right)
	+ \frac{2 \mu q_e^3 r^2}{(r^3 +q_e^3)^{7/3}} \Big], 
	\label{scem}
\end{eqnarray}
and the effective potential corresponding to the modes \textbf{II} 
of the electromagnetic perturbations has the following form 
\cite{Crispino:2000jx,LopezOrtega:2006vn}
\begin{eqnarray}
	V_2(r) &=& \left(1 -\frac{\mu r^2}{(r^3 +q_e^3)^{4/3}}\right)\,
	\Big[\frac{(l+1)^2}{r^2} \notag\\
	&&- \frac{3}{4r^2}\; 
	\left(1 - \frac{\mu r^2}{(r^3 +q_e^3)^{4/3}}\right)
	- \frac{2 \mu q_e^3 r^2}{(r^3 +q_e^3)^{7/3}} \Big]. 
	\label{vecem}
\end{eqnarray}
To see the nature, we illustrate these expressions of the effective 
potential functions versus radius $r$. One can see the typical 
behavior of these effective potentials in Figs.~\ref{semp} and 
\ref{vemp}. We find that an increase in the magnitude of electric 
charge $q_e$ changes the height of the potential barrier. Similarly, 
the height of the potential barrier also varies with an increase in 
multipole number $l$.
\begin{table}[h]
	\caption{ \label{table2} Real and imaginary parts of the QNMs 
	frequencies in the case of modes \textbf{I} of the electromagnetic 
	perturbations ($\mu=1$).}
	\centering
	\begin{tabular}{|l|l|l|l|l|}
		\hline
	\multicolumn{1}{|c|}{  } &  \multicolumn{1}{c|}{  $l=1, n=0$ } & 
	\multicolumn{1}{c|}{  $l=2, n=0$ } & \multicolumn{1}{c|}{ $l=2, n=1$}\\
		\hline
	$q_e$ & $\omega \,(WKB)$ & $\omega \,(WKB)$ & $\omega \,(WKB)$ \\ 
	\hline
	0 	& 0.73685 - 0.31537 i & 1.34123 - 0.33725 i & 1.21139 - 1.04385 i  \\
	0.1 & 0.73771 - 0.31492 i & 1.34165 - 0.33712 i & 1.21210 - 1.04342 i  \\
	0.2 & 0.74378 - 0.31183 i & 1.34453 - 0.33617 i & 1.21702 - 1.04039 i  \\
	0.3 & 0.76020 - 0.30411 i & 1.35250 - 0.33352 i & 1.23035 - 1.03188 i  \\
	0.4 & 0.79081 - 0.29230 i & 1.36856 - 0.32783 i & 1.25617 - 1.01378 i  \\
	0.5 & 0.82919 - 0.28502 i & 1.39710 - 0.31630 i & 1.29788 - 0.97684 i  \\ 
	\hline
	\end{tabular}
\end{table}
\begin{table}[h]
	\caption{\label{table3} Real and imaginary parts of the QNMs 
	frequencies in the case of modes \textbf{II} of the electromagnetic 
	perturbations ($\mu=1$).}
	\centering
	\begin{tabular}{|l|l|l|l|l|}
		\hline
	\multicolumn{1}{|c|}{  } &  \multicolumn{1}{c|}{  $l=1, n=0$ } & 
	\multicolumn{1}{c|}{  $l=2, n=0$ } & \multicolumn{1}{c|}{ $l=2, n=1$}\\
		\hline
	$q_e$ & $\omega \,(WKB)$ & $\omega \,(WKB)$ & $\omega \,(WKB)$  \\ 
	\hline
	0 	& 0.95143 - 0.35304 i & 1.46852 - 0.35248 i & 1.34827 - 1.08979 i \\
	0.1 & 0.95174 - 0.35296 i & 1.46894 - 0.35234 i & 1.34892 - 1.08934 i \\
	0.2 & 0.95393 - 0.35240 i & 1.47186 - 0.35137 i & 1.35348 - 1.08614 i \\
	0.3 & 0.95996 - 0.35068 i & 1.47993 - 0.34858 i & 1.36596 - 1.07699 i \\
	0.4 & 0.97226 - 0.34634 i & 1.49638 - 0.34253 i & 1.39060 - 1.05692 i \\
	0.5 & 0.99438 - 0.33563 i & 1.52601 - 0.33011 i & 1.43181 - 1.01517 i \\
		\hline
	\end{tabular}
\end{table}
\begin{figure*}
	\begin{center}
		\includegraphics[width=7.9cm]{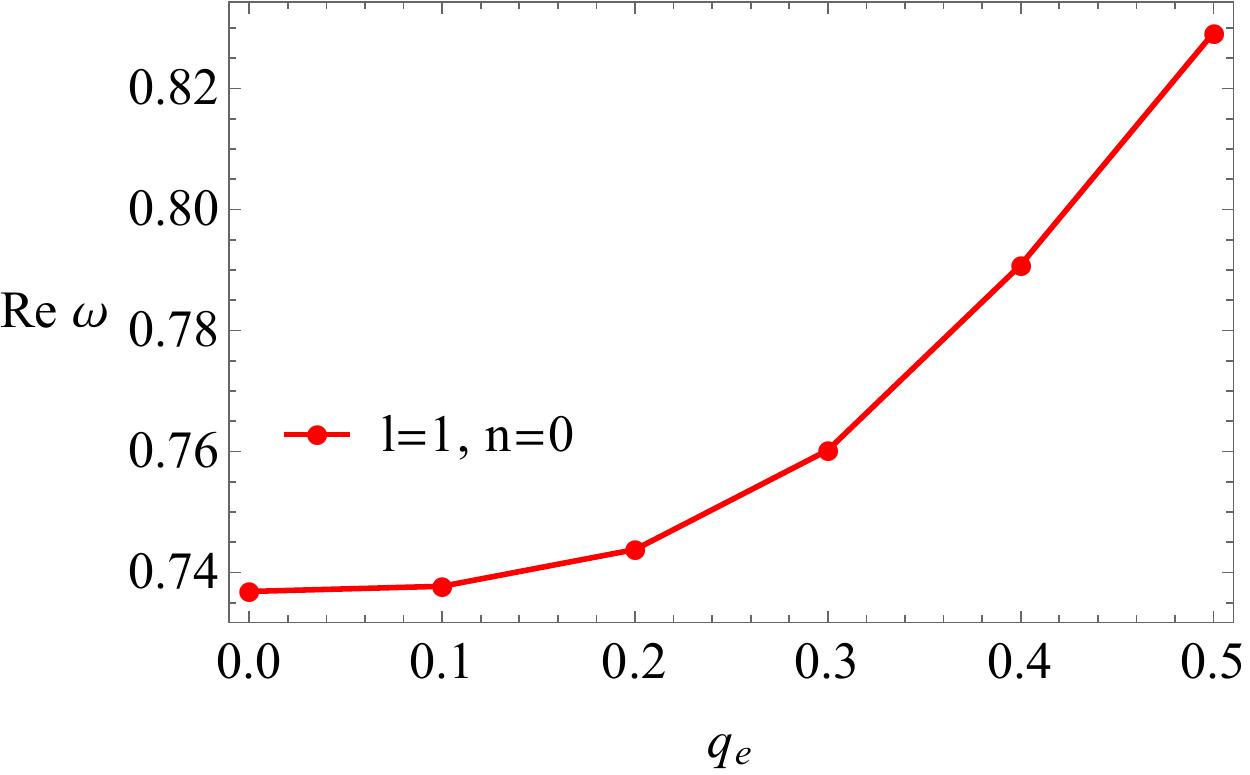}
		\includegraphics[width=7.9cm]{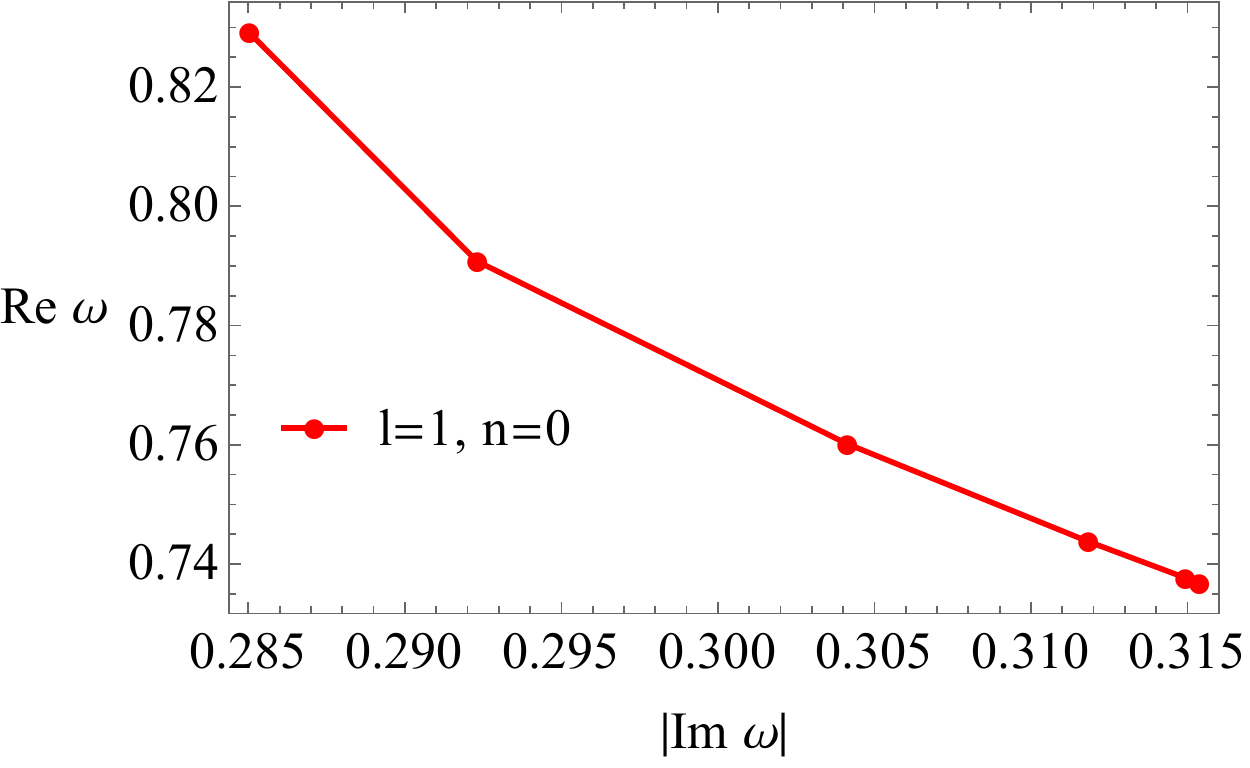}
		\includegraphics[width=7.9cm]{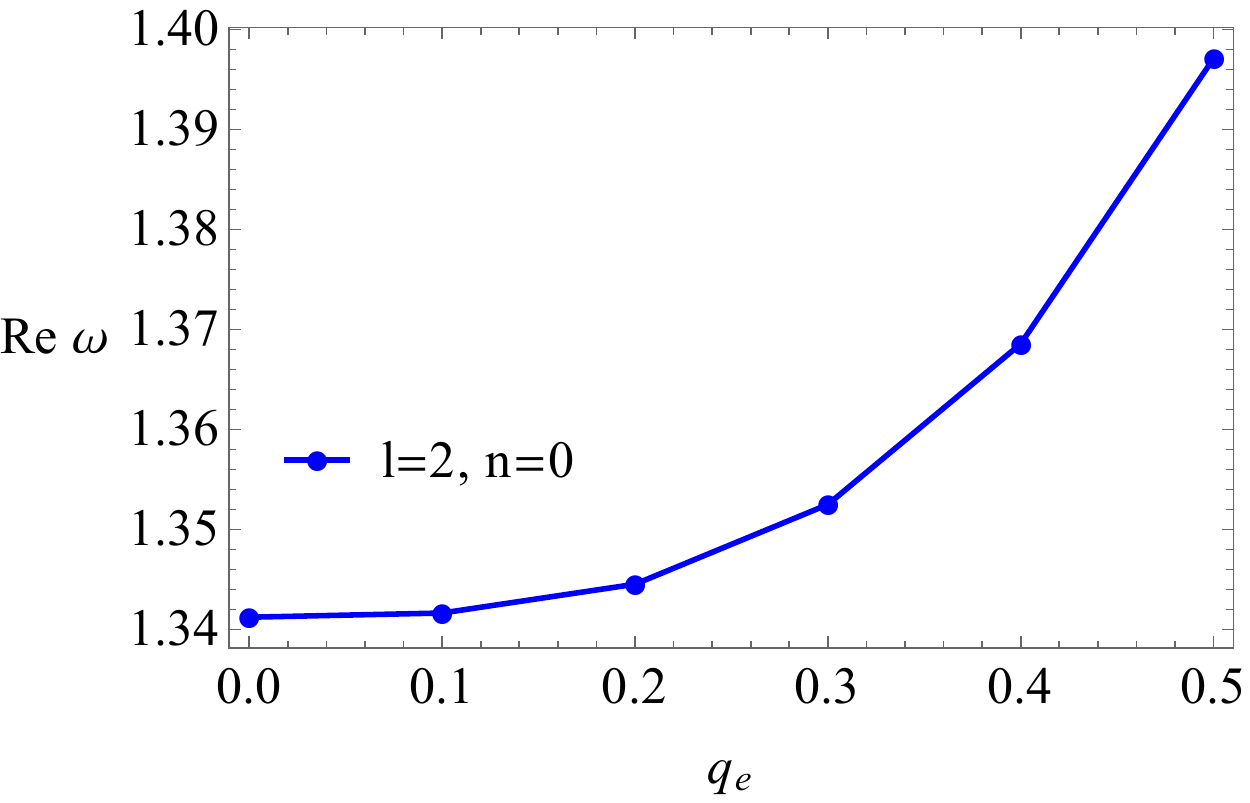}
		\includegraphics[width=7.9cm]{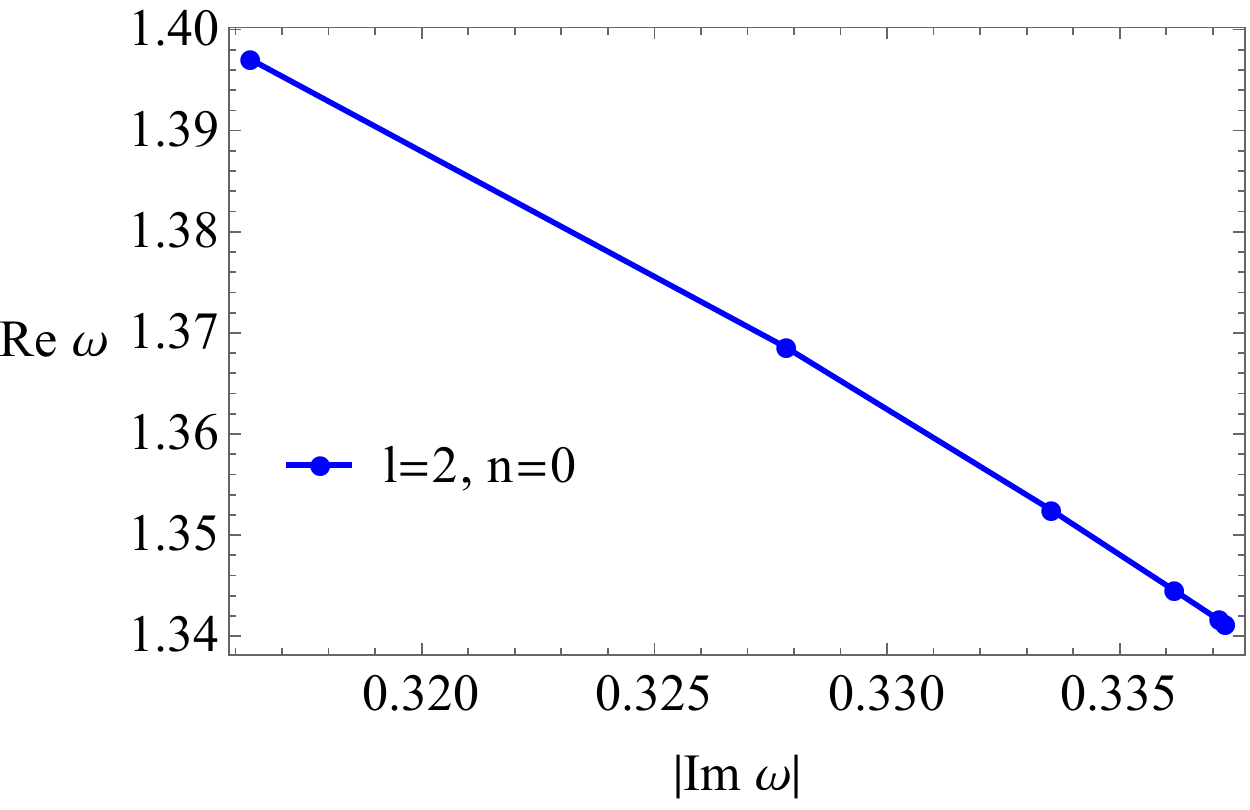}
		\includegraphics[width=7.9cm]{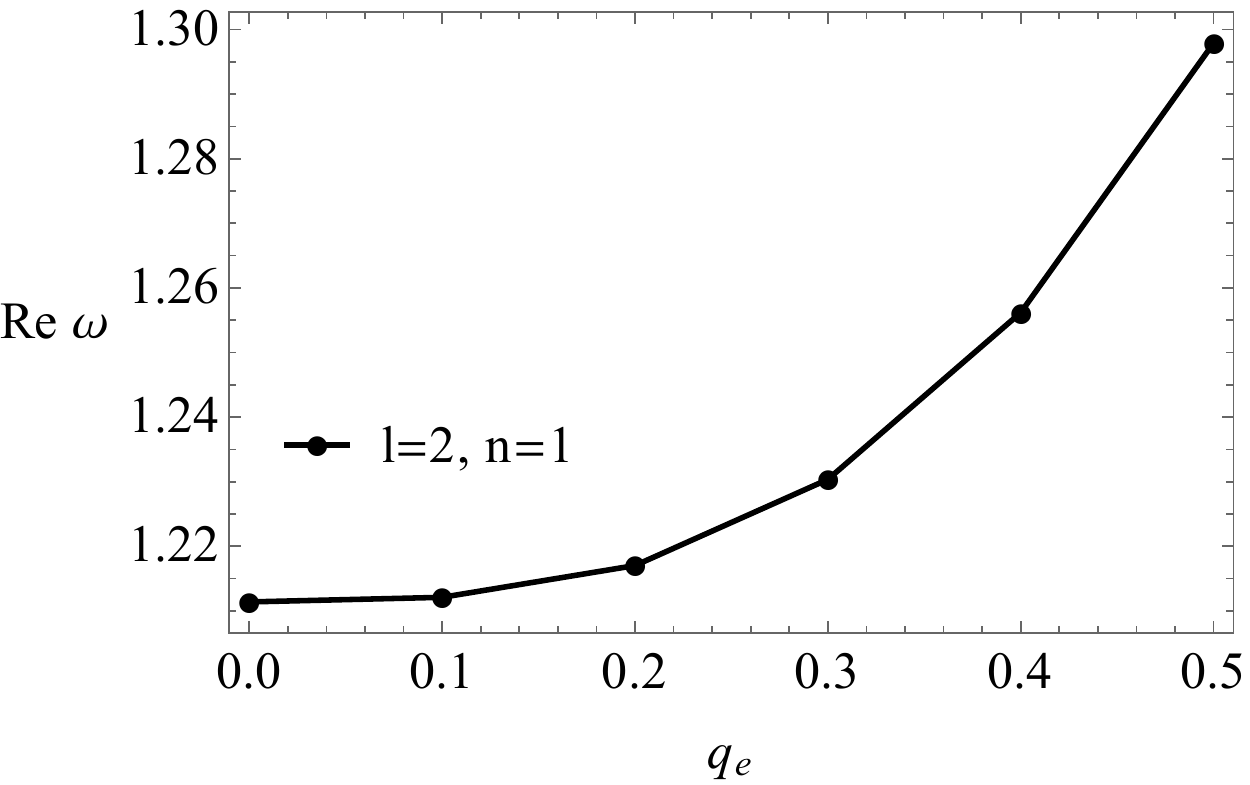}
		\includegraphics[width=7.9cm]{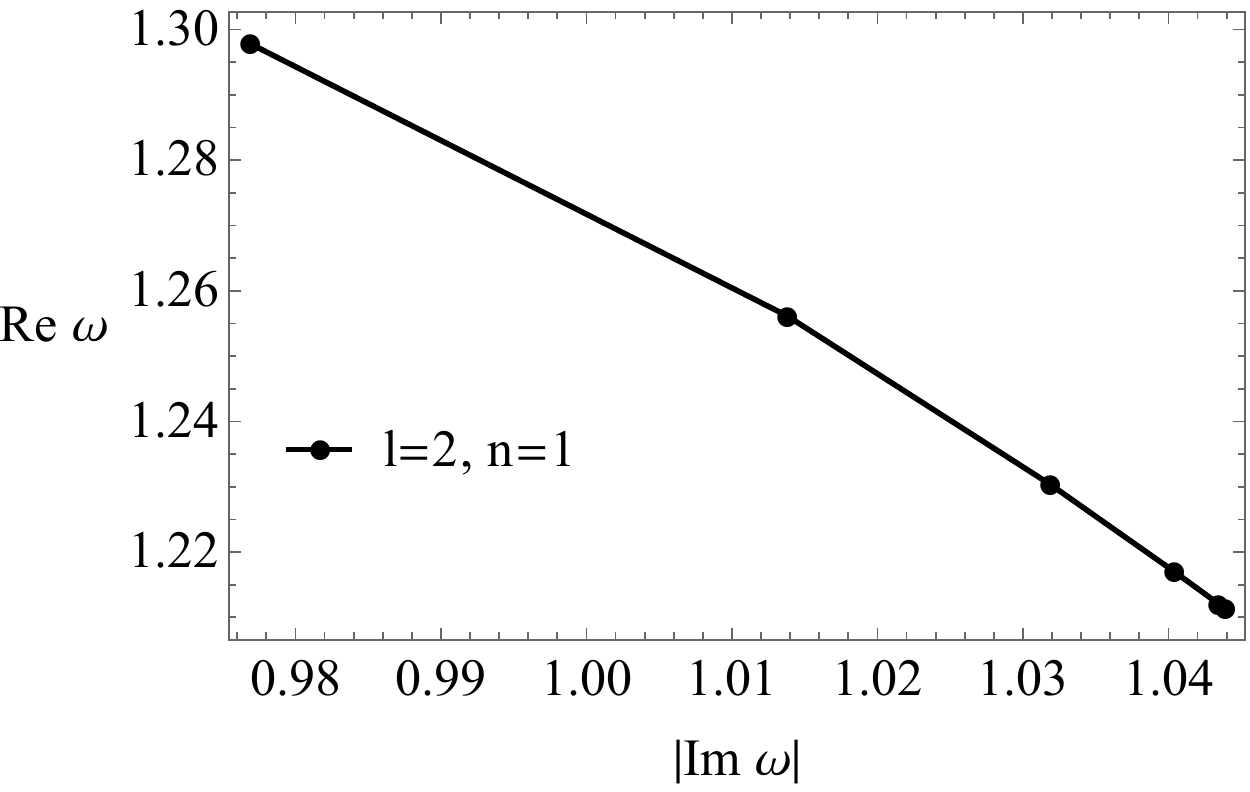}
	\caption{\label{sqnemp1} (Left panel) Plot showing the dependence of 
	real part of the QNMs with electric charge $q_e$ for the modes 
	\textbf{I} of the electromagnetic field perturbations. (Right panel) 
	Plot showing the dependence of real part of the QNMs versus the 
	imaginary part of the QNMs in absolute value ($\mu=1$).}
	\end{center}
\end{figure*}
\begin{figure*}
	\includegraphics[width=7.9cm]{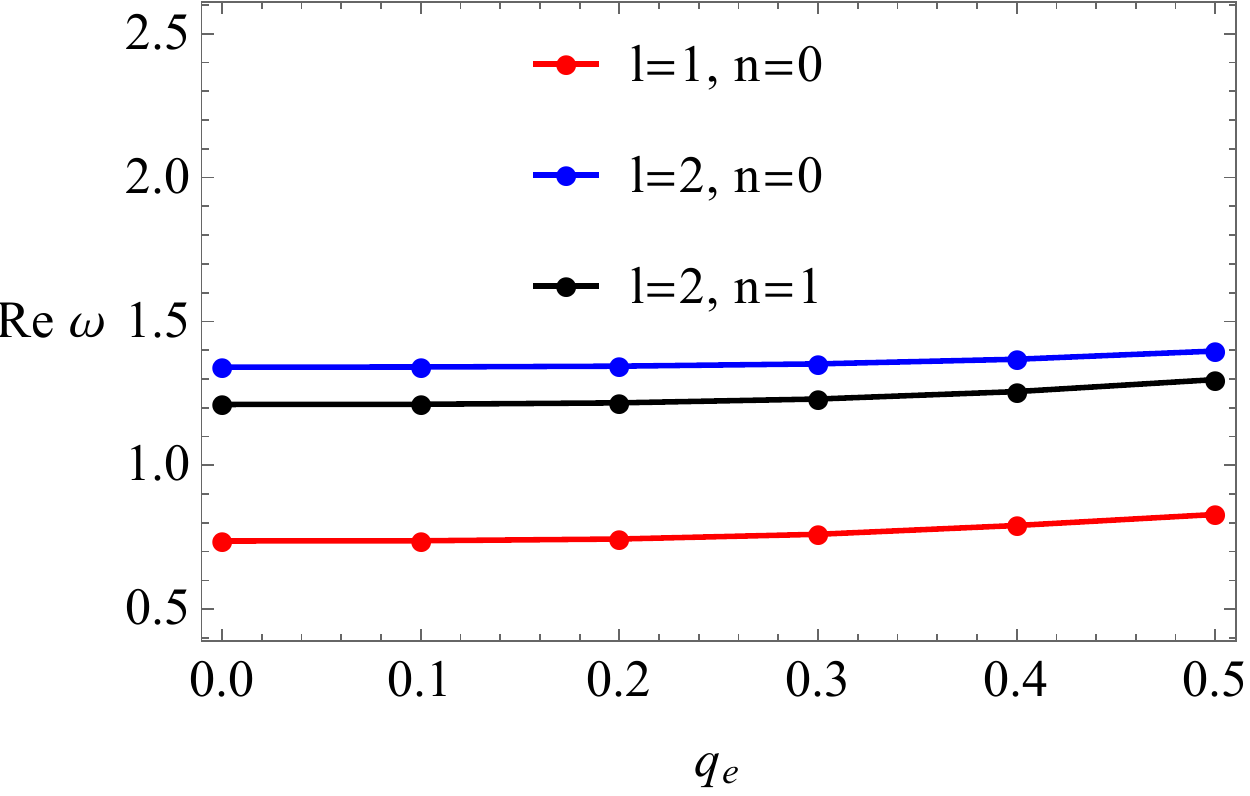}
	\includegraphics[width=7.9cm]{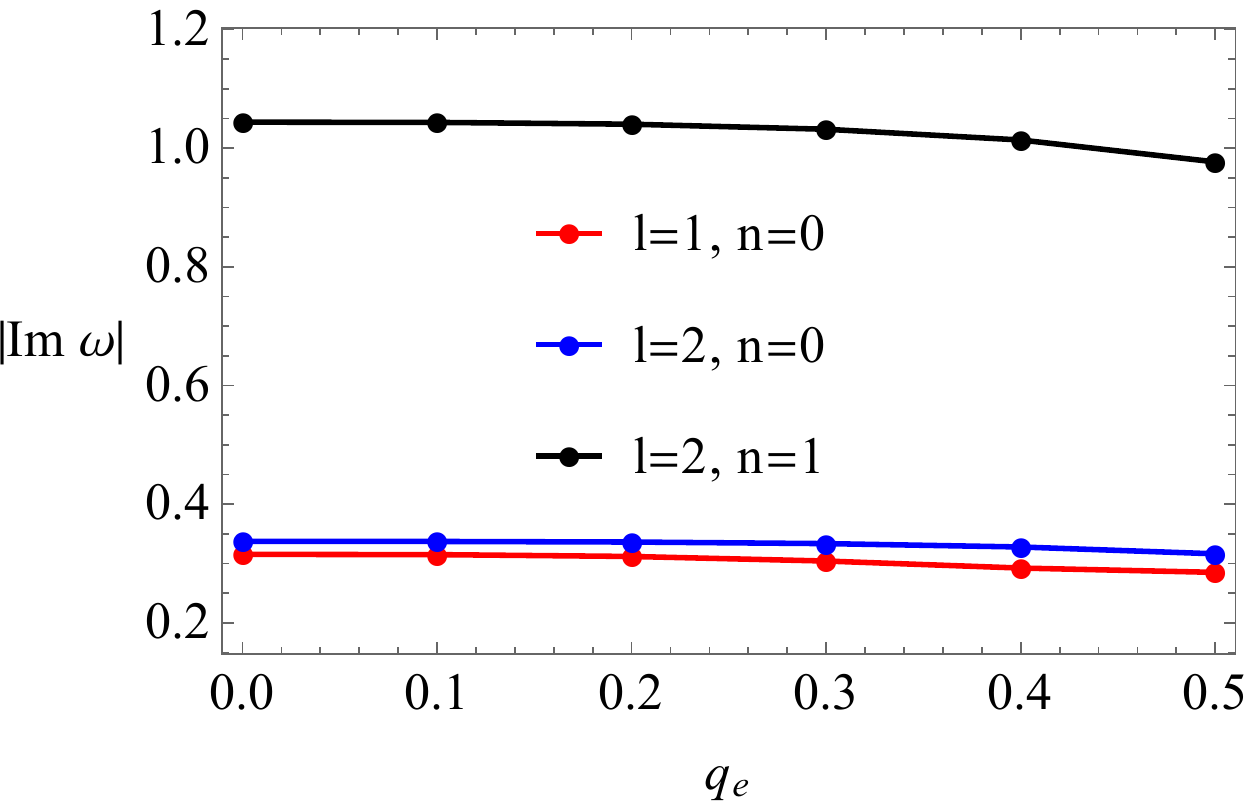}
	\caption{\label{sqnemp2} (Left panel) Plots showing the dependence of 
	real part of the QNMs with electric charge $q_e$ for different values 
	of $l$ and $n$ for the modes \textbf{I} of the electromagnetic 
	perturbations. 
	(Right panel) Plots showing the dependence of real part of the QNMs 
	in absolute value with electric charge $q_e$ for different values 
	of $l$ and $n$ ($\mu=1$). }
\end{figure*}
\begin{figure*}
	\begin{center}
		\includegraphics[width=7.9cm]{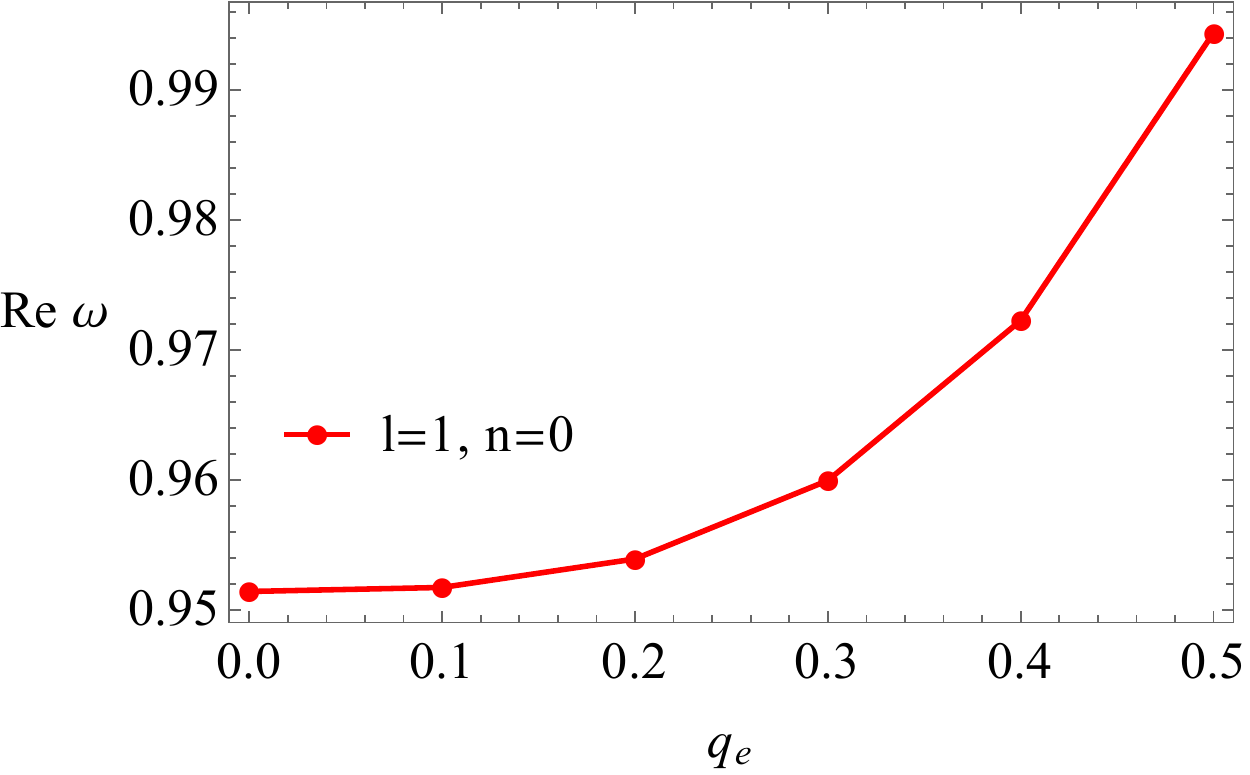}
		\includegraphics[width=7.9cm]{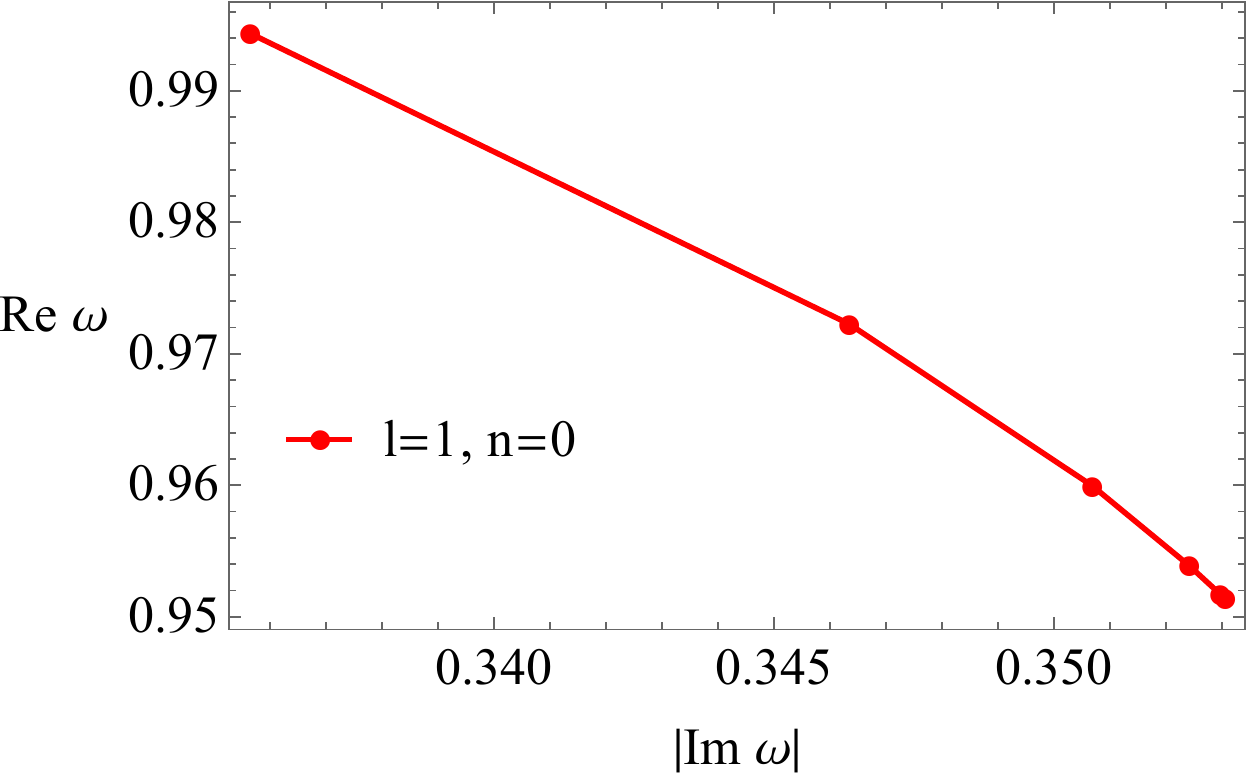}
		\includegraphics[width=7.9cm]{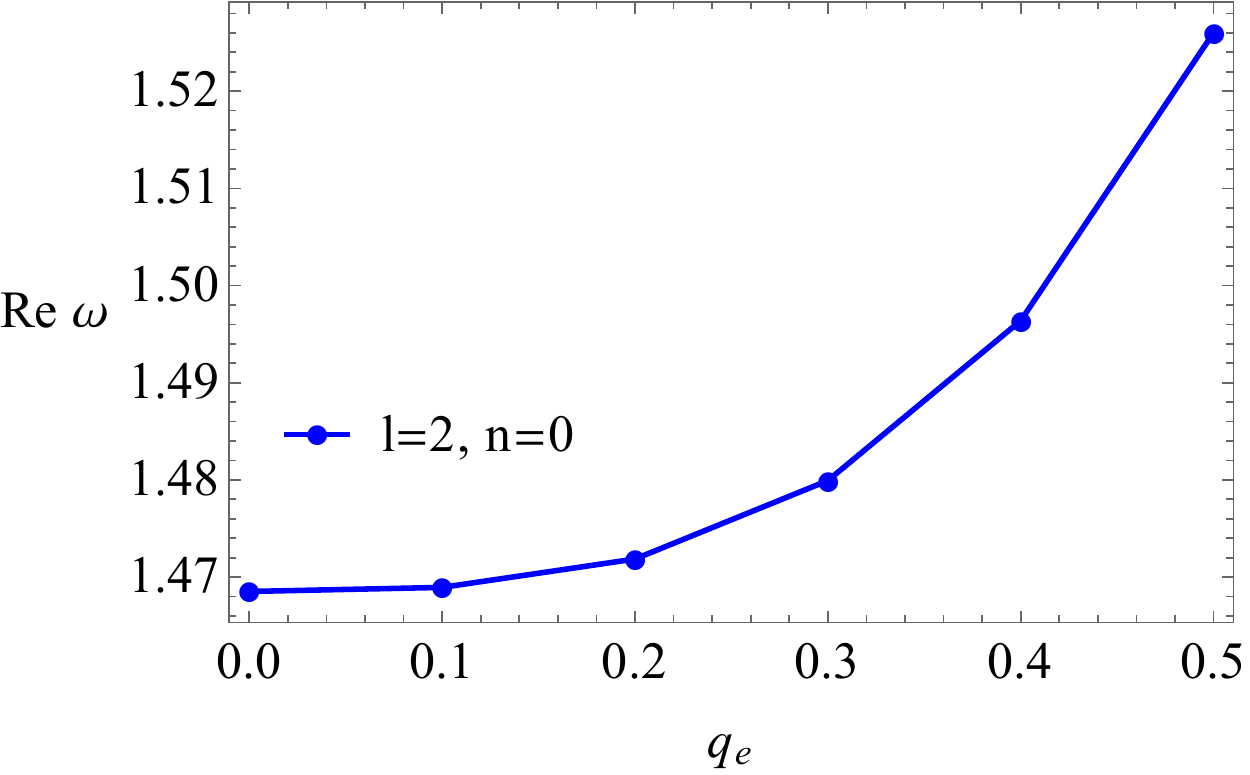}
		\includegraphics[width=7.9cm]{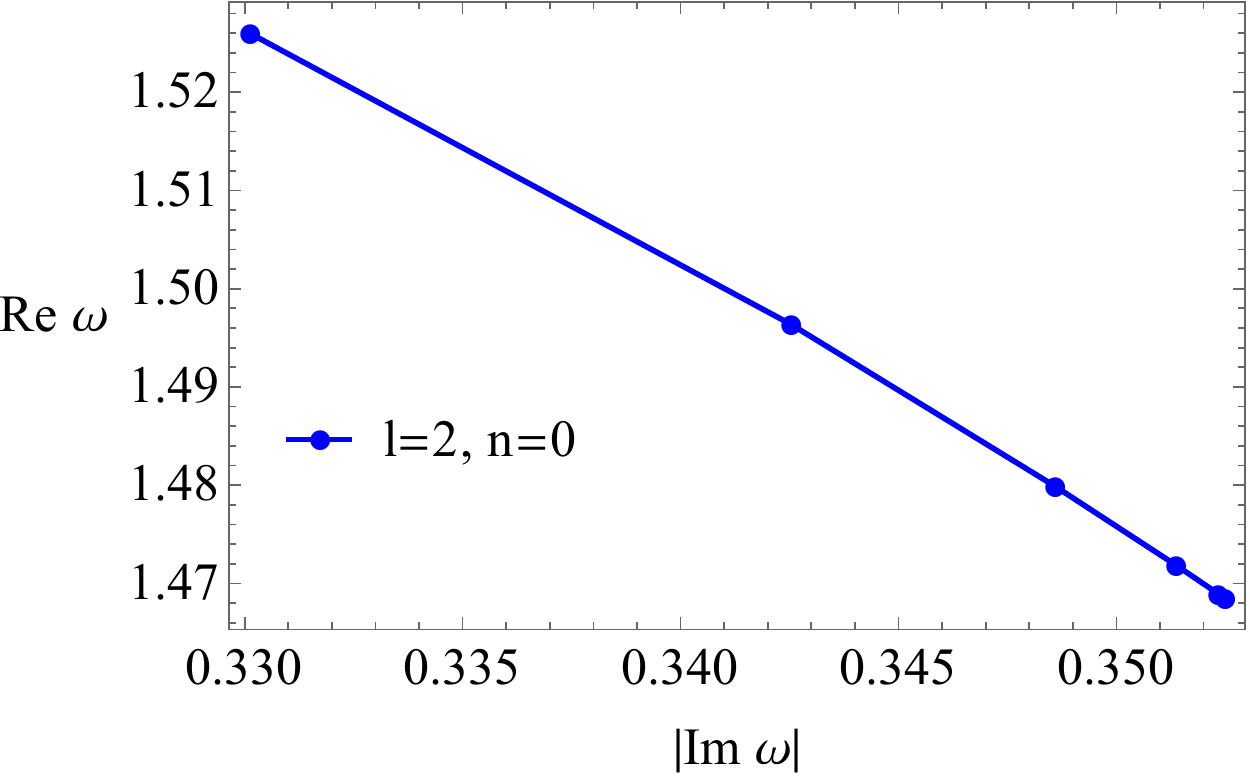}
		\includegraphics[width=7.9cm]{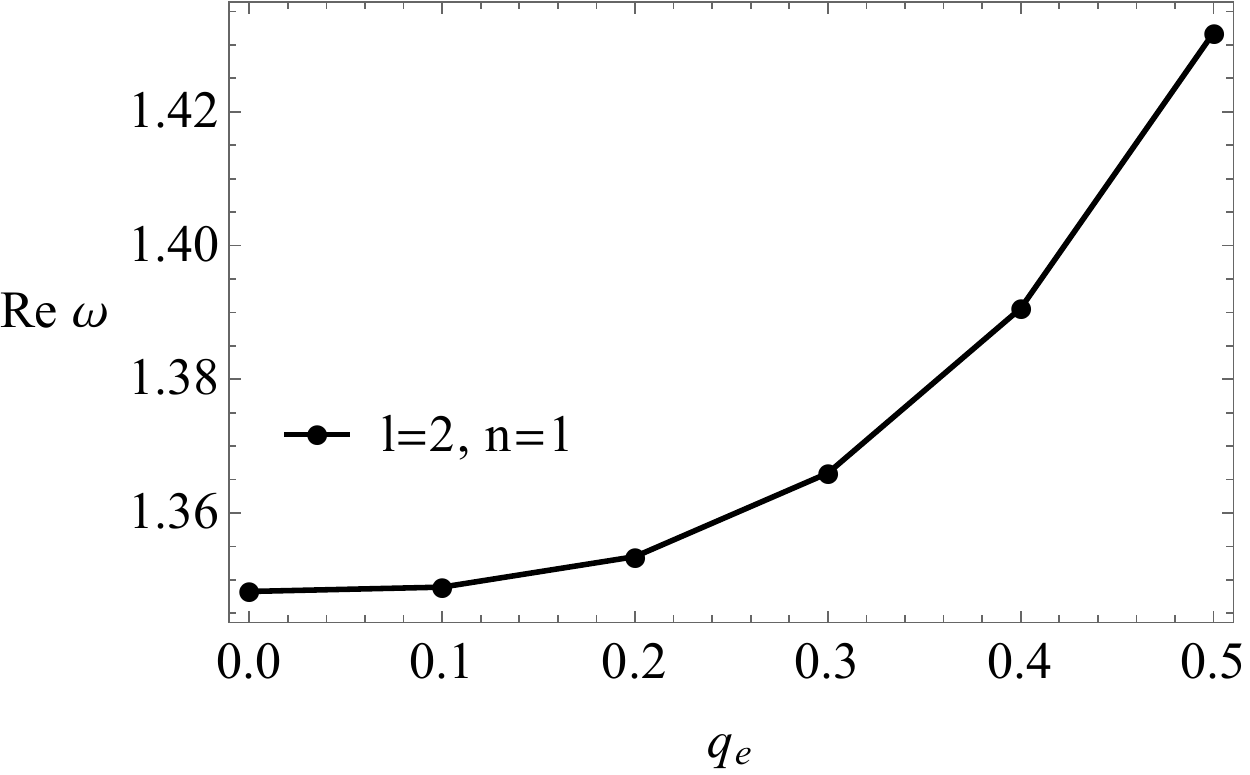}
		\includegraphics[width=7.9cm]{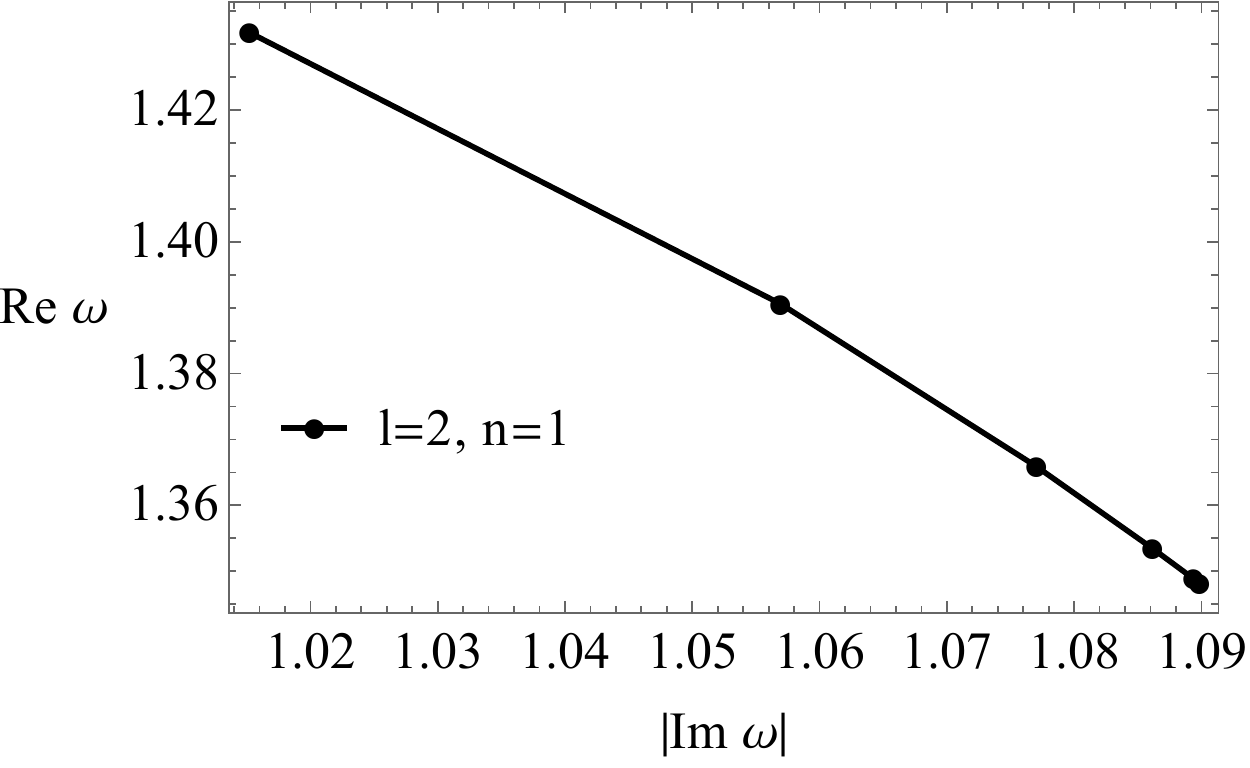}
	\caption{\label{vqnemp1} (Left panel) Plots showing the dependence 
	of real part of the QNMs with electric charge $q_e$ for the	modes 
	\textbf{II of the} electromagnetic field perturbations. 
	(Right panel) Plots showing the dependence of real part of the QNMs 
	versus the imaginary part of the QNMs in absolute value ($\mu=1$).}
	\end{center}
\end{figure*}
\begin{figure*}
	\includegraphics[width=7.9cm]{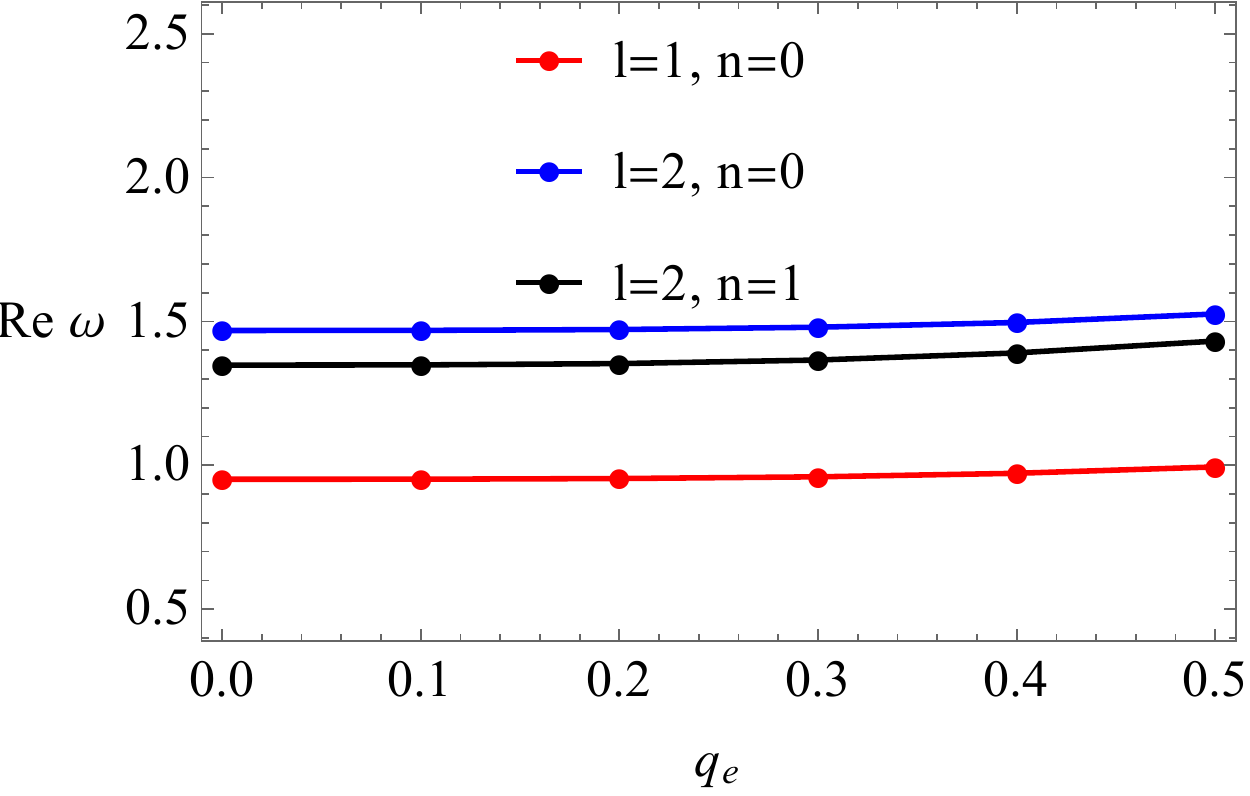}
	\includegraphics[width=7.9cm]{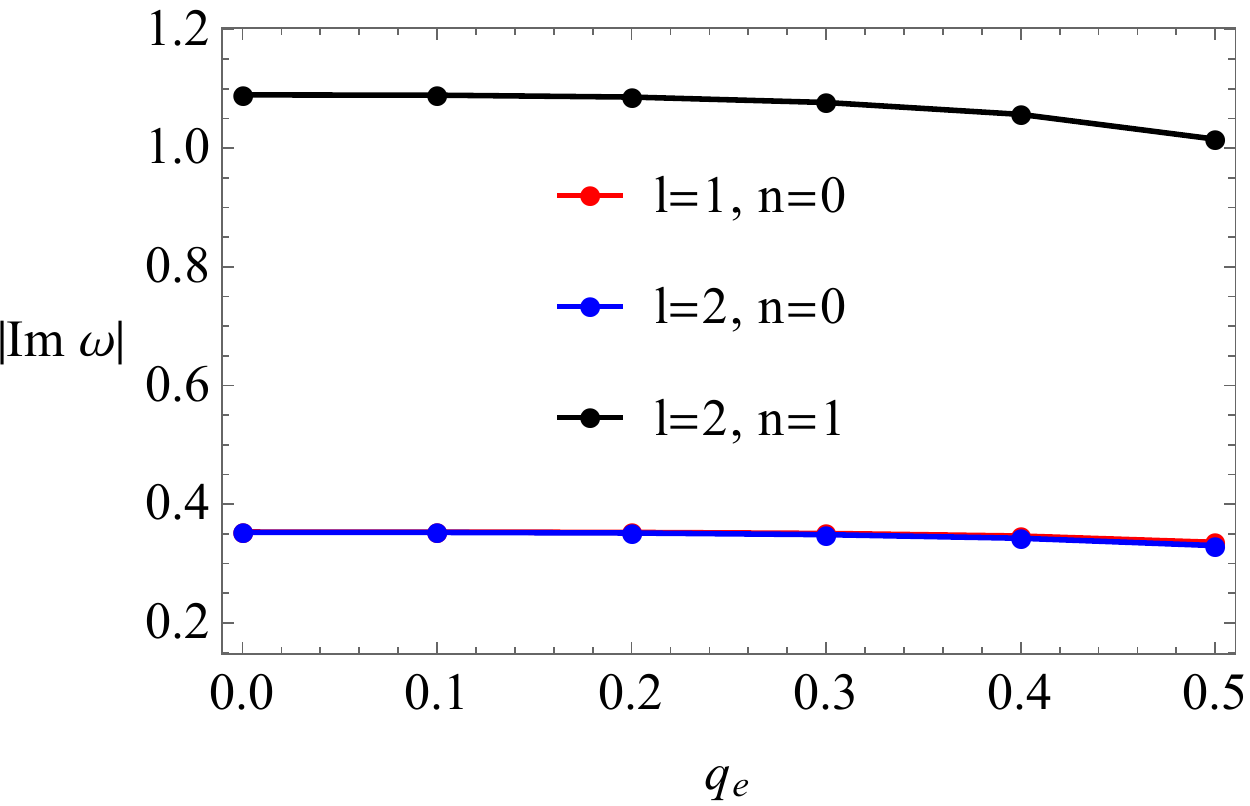}
	\caption{\label{vqnemp2} (Left panel) Plots showing the dependence 
	of real part of the QNMs with electric charge $q_e$ for different 
	values of $l$ and $n$ for the modes \textbf{I} of the electromagnetic 
	perturbations. 
	(Right panel) Plots showing the dependence of real part of the QNMs 
	in absolute value with electric charge $q_e$ for different values of 
	$l$ and $n$ ($\mu=1$). }
\end{figure*}
We further present numerical calculations of the QNMs frequencies 
for the corresponding modes \textbf{I} and modes \textbf{II} of the 
electromagnetic field perturbations (cf. Table~\ref{table2} and 
\ref{table3}). We observe that by increasing the electric charge 
$q_e$, the real part of the QNMs frequencies increases while the 
imaginary part decreases in absolute value compared to the 
Schwarzschild-Tangherlini black hole. This particular effect can be 
clearly seen from Figs.~\ref{sqnemp1}, \ref{sqnemp2}, \ref{vqnemp1}, 
and \ref{vqnemp2}. Moreover, we analyze the numerical results of the 
real/imaginary part in absolute value of the QNMs frequencies for the 
corresponding modes \textbf{I} and modes \textbf{II} of the electromagnetic 
field perturbations. We find that the modes \textbf{II} of the electromagnetic 
perturbations oscillate and damp more rapidly in comparison to the modes 
\textbf{I} (cf. Table~\ref{table2} and \ref{table3}).

Furthermore, we compare the numerical results of the 
scalar field perturbations and the electromagnetic field perturbations.
We find that the numerical values of the real/imaginary part of the QNMs 
frequencies in absolute value are comparatively higher in the scalar 
field perturbations (cf. Table~\ref{table1}, \ref{table2} and \ref{table3}). 
It turns out that the scalar field perturbations oscillate more rapidly 
compared to the electromagnetic ones. On the other hand, the scalar 
field perturbations damp more rapidly than the electromagnetic ones. 
In general, both the electromagnetic and scalar field perturbations 
in presence of the electric charge $q_e$ decay more slowly compared 
to the Schwarzschild-Tangherlini black holes.

\section{Scattering and greybody factors}
\label{greybf}
Now we are going to discuss the greybody factors in 5D electrically 
charged Bardeen black holes spacetime. This study is vital to determine, 
for instance, the amount of initial quantum radiation in the vicinity 
of the event horizon of black hole, which is reflected back to it by the 
potential barrier. Therefore, it is natural to interpret the greybody 
factors as the tunneling probability of the wave through the barrier 
determined by the effective potential in given black hole spacetime. 
We begin with the Schr{\"o}dinger-like equation that describes the 
scattering of waves in 5D electrically charged Bardeen black hole 
spacetime. The asymptotic solutions of the Schr{\"o}dinger-like 
equation are given as follows
\begin{eqnarray}
    \Psi &=& A e^{-i \omega r_{\star}} +B e^{i \omega r_{\star}},
    \quad  r_{\star} \to -\infty, \\ [2mm]
    \Psi &=& C e^{-i \omega r_{\star}} +D e^{i \omega r_{\star}},
    \quad   r_{\star} \to +\infty ,
\end{eqnarray}
where $A, B, C$, and $D$ are functions of the frequency $\omega$. 
Furthermore, we need to impose the conditions $B(\omega)= 0$, for 
waves coming to the black hole from infinity and 
$R(\omega) = D(\omega)/C(\omega)$, for the reflection amplitude. 
It is worth noting that these waves are identical to that of the 
scattered ones due to the black holes' event horizon. The 
transmission amplitude is given by $T(\omega)= A(\omega)/C(\omega)$, 
thence
\begin{eqnarray}
	\Psi &=& T e^{-i \omega r_{\star}} \quad r_{\star} \to -\infty, 
	\\ [2mm]
	\Psi &=& e^{-i \omega r_{\star}} +R e^{i \omega r_{\star}},
	\quad r_{\star} \to +\infty.
\end{eqnarray}
Our next task is to determine the square of the wave function's 
amplitude. To achieve this goal, we can use the fact that the total
probability of finding wave must obey the normalization condition 
given by
\begin{equation}
	|R|^2+|T|^2=1. \label{ampcond}
\end{equation}
\begin{figure*}
    \begin{center}
        \includegraphics[width=8 cm]{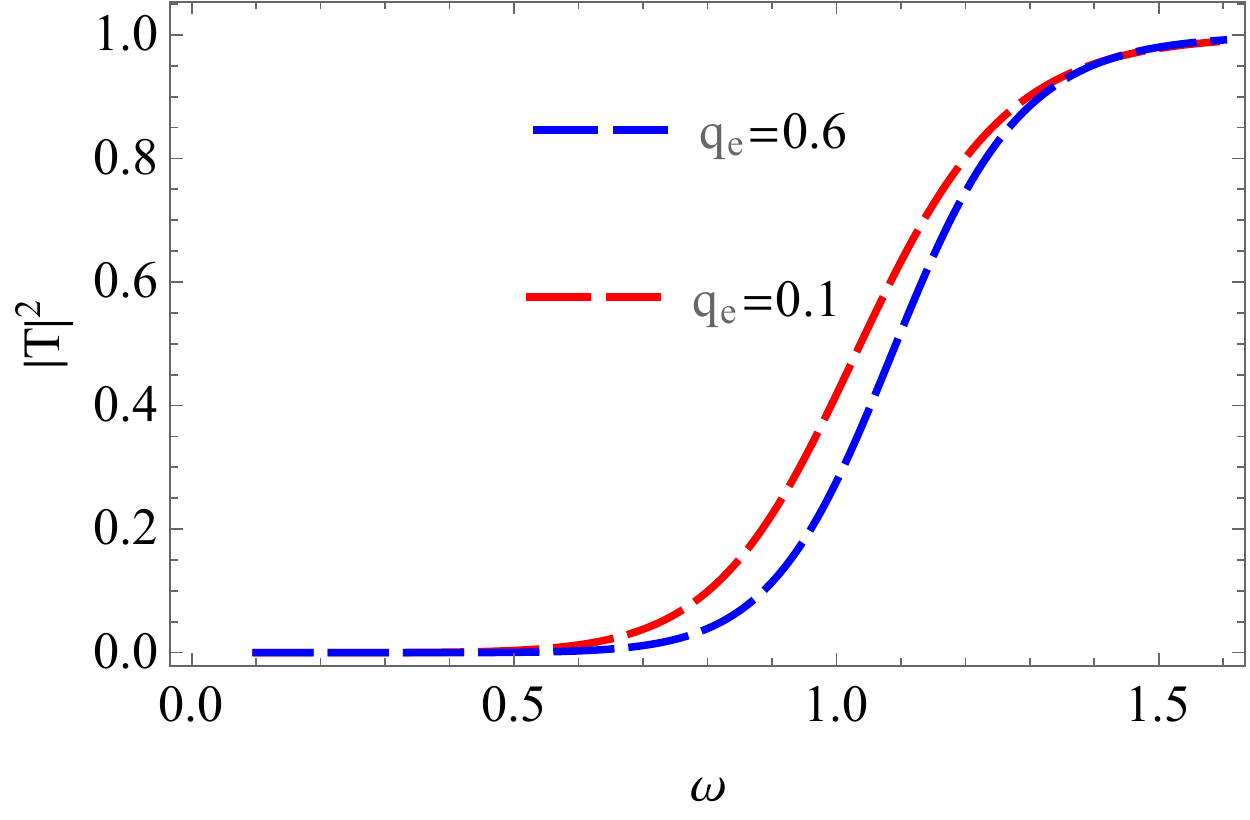}
        \includegraphics[width=8 cm]{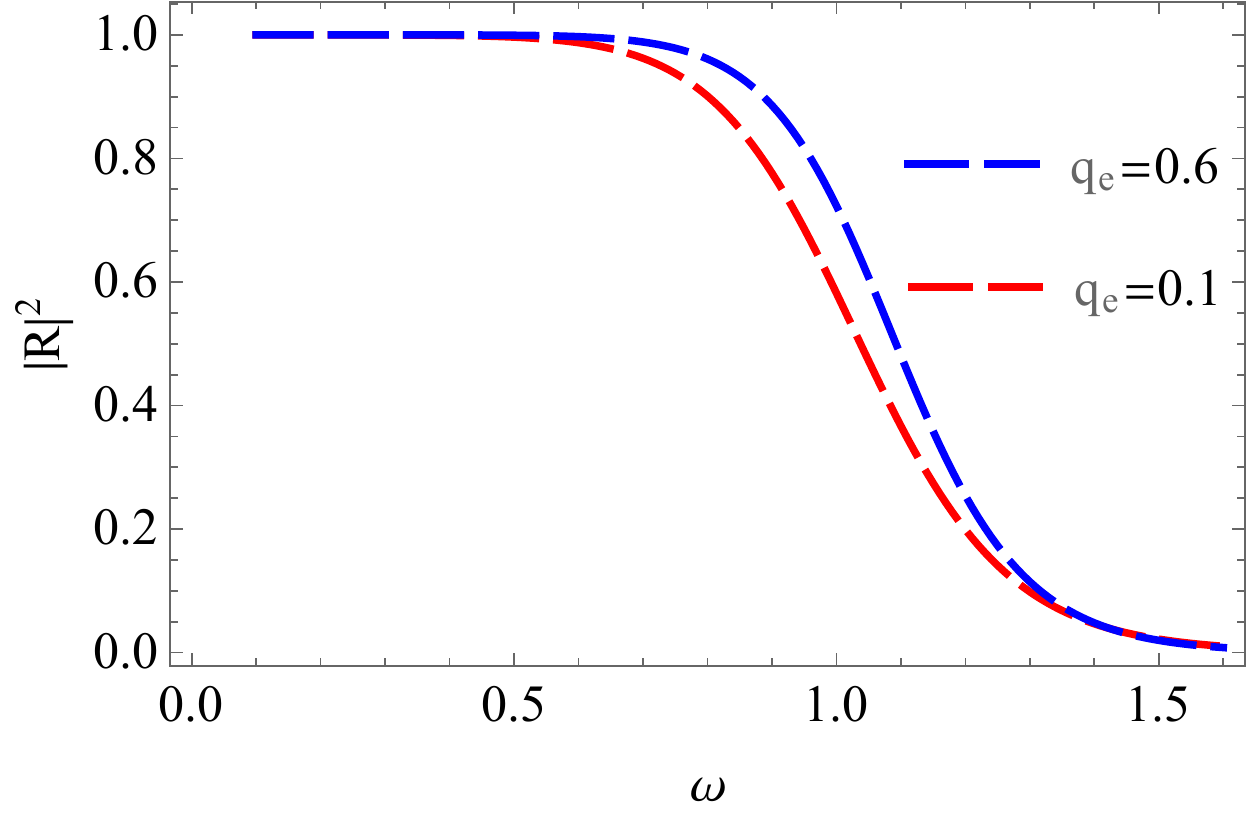}
        \includegraphics[width=8 cm]{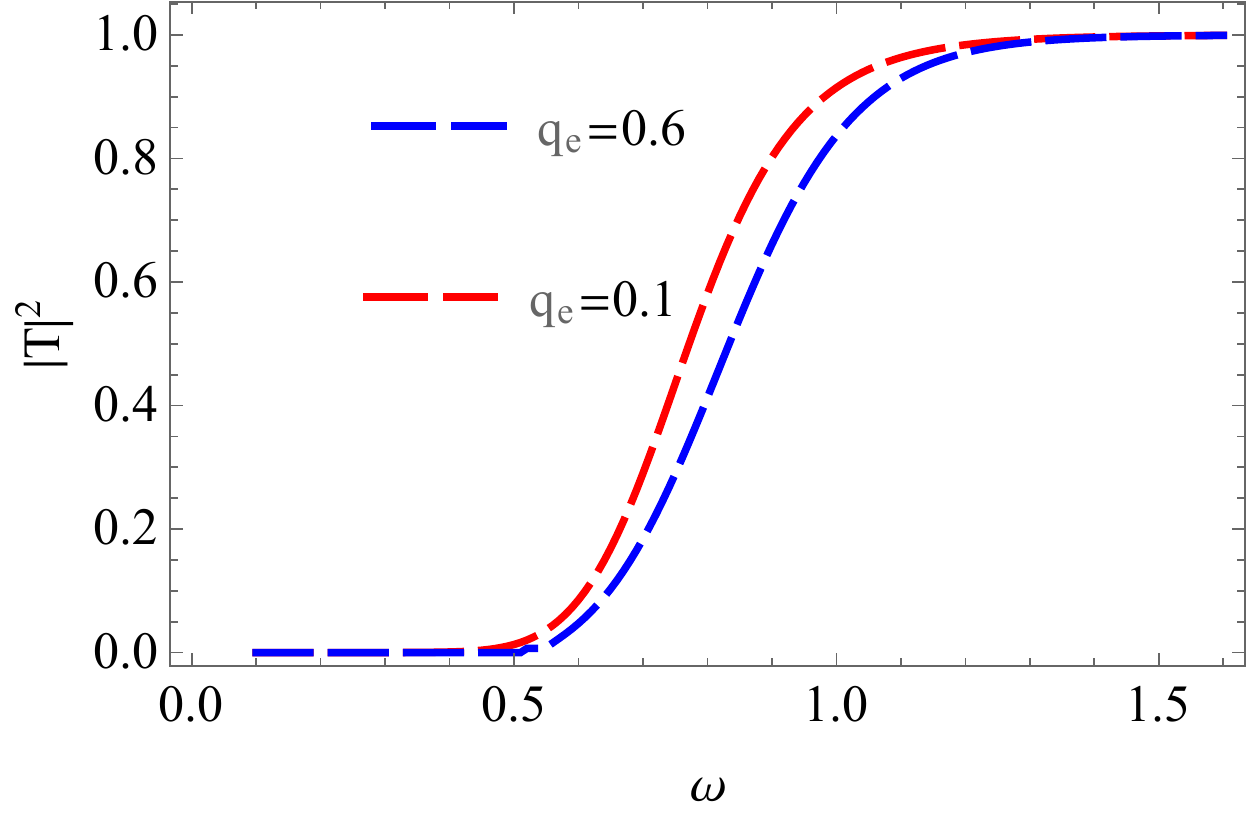}
        \includegraphics[width=8 cm]{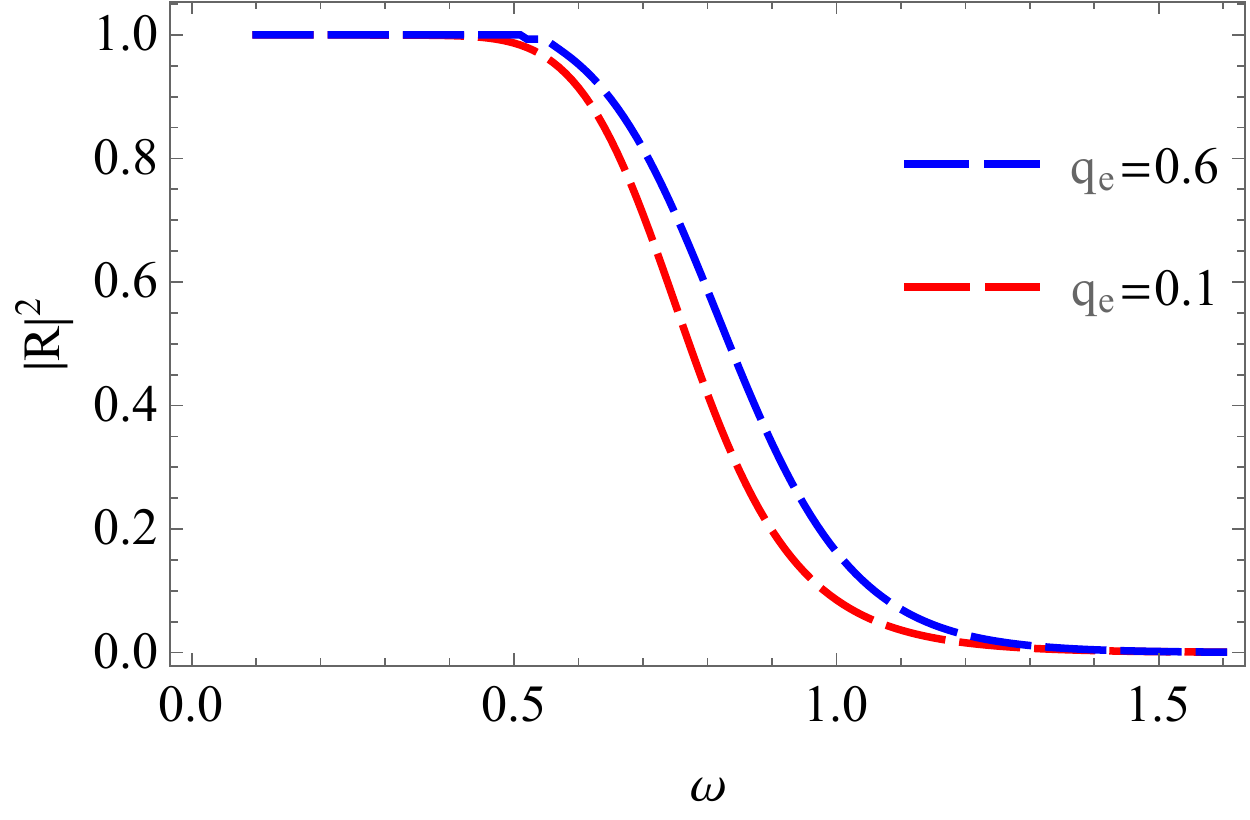}
        \includegraphics[width=8 cm]{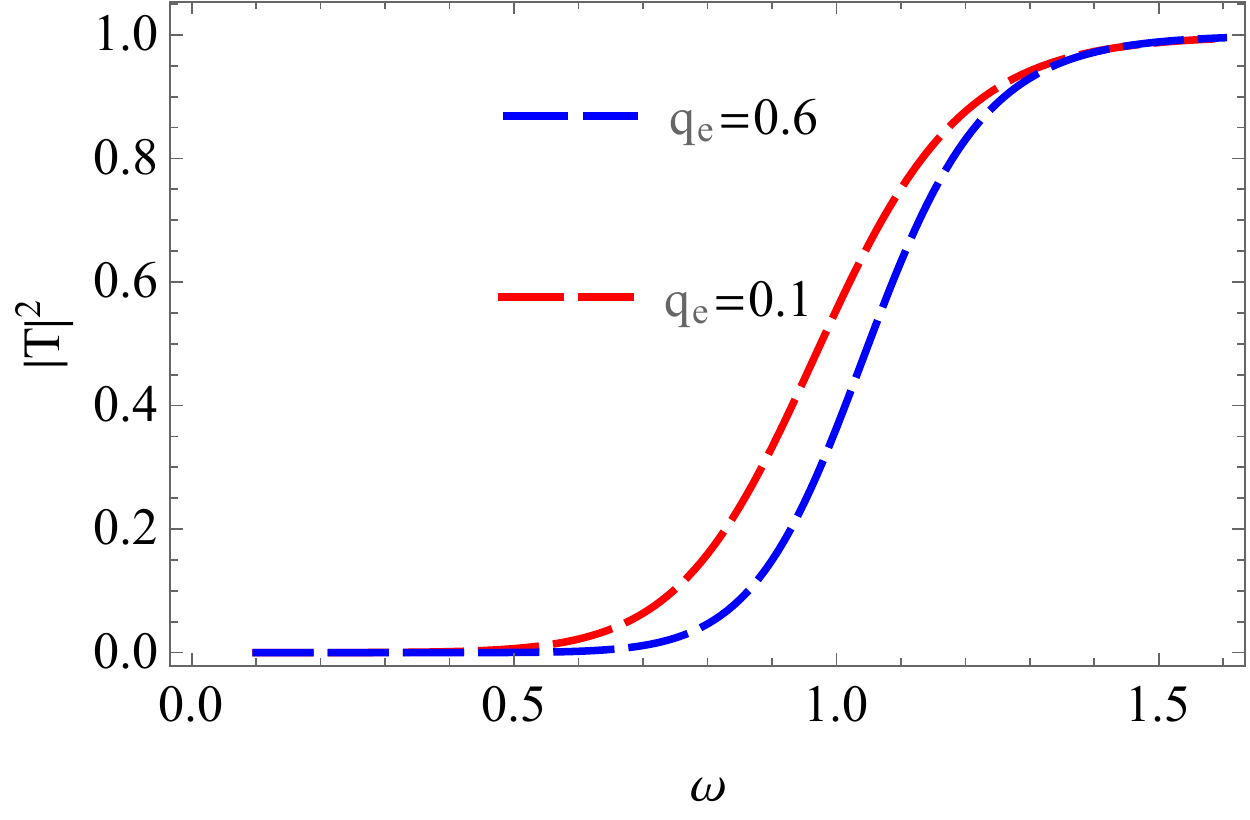}
        \includegraphics[width=8 cm]{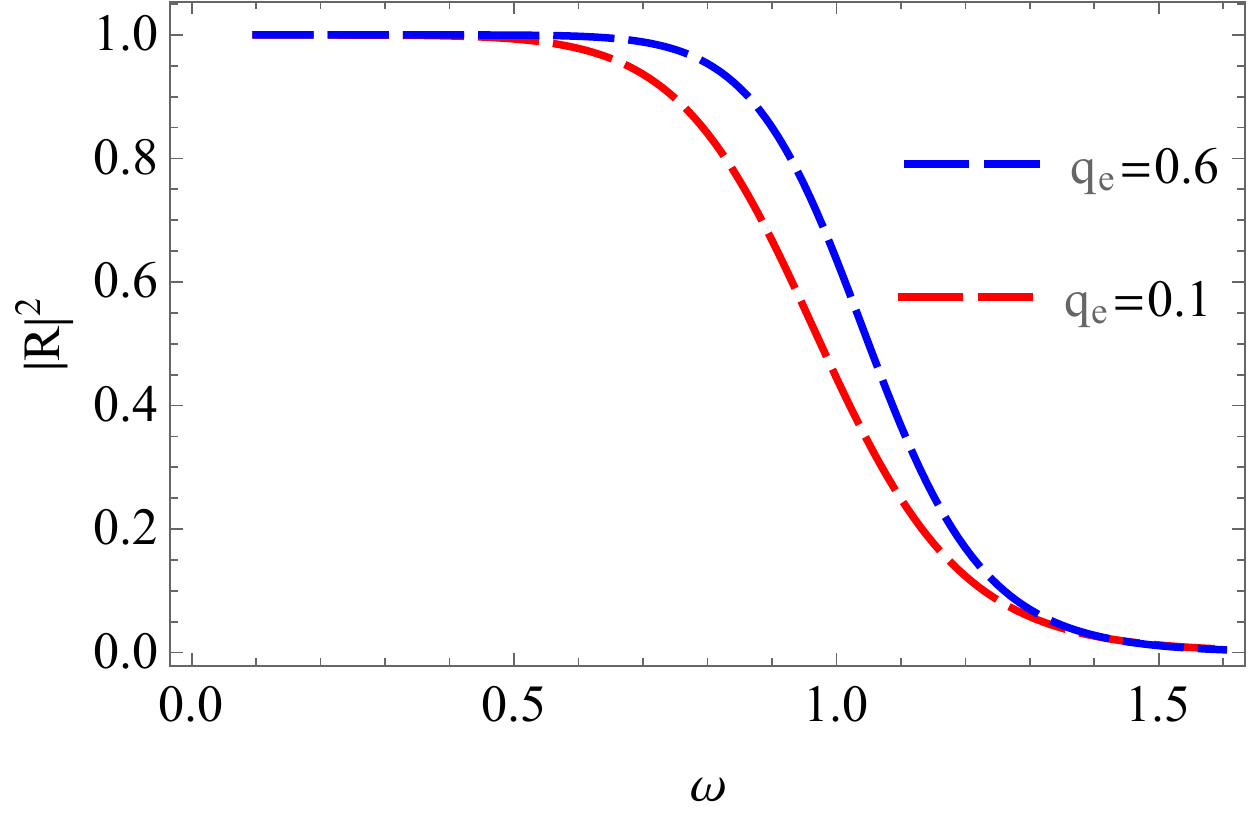}
    \caption{\label{rtcof} (Left panel): From top to bottom, plots 
    showing the behavior of transmission coefficients for the scalar 
    field perturbations and electromagnetic field perturbations for 
    the modes \textbf{I} and \textbf{II}, respectively. 
    (Right panel): From top to bottom, plots showing the reflection 
    coefficient for the scalar field perturbations and electromagnetic 
    field perturbations for the modes \textbf{I} and \textbf{II}, 
    respectively ($\mu=1$ and $l=1$).}
    \end{center}
\end{figure*}
While discussing the greybody factors, we consider the scenario 
$\omega \simeq V(r_0)$, which is the most compelling case. Note 
that, in general, the Schr{\"o}dinger-like equation cannot be solved 
analytically. Therefore, to compute the reflection and transmission 
coefficients, we follow the WKB approximation approach. On using the 
above equations, it is not difficult to show that the reflection 
coefficient is given by
\begin{equation}
	R=\frac{1}{\sqrt{1+\exp(-2 \pi i K)}},\label{refamp}
\end{equation}
where $K$ is expressed as follows
\begin{equation}
	K=i\frac{\omega_n^2-V(r_0)}{\sqrt{-2\,V''(r_0)}}
	-\sum_{i=2}^{6}\Lambda_i.
\end{equation}
On having the expression of the reflection coefficient, now we can 
use the condition \eqref{ampcond} to obtain the transmission 
coefficient which takes the following form
\begin{equation}
	|T|^2= 1 -\left|\frac{1}{\sqrt{1+\exp(-2 \pi i K)}}\right|^2.
\end{equation}
To see the nature of these coefficients, we plot them against frequency 
$\omega$ for different values of the electric charge $q_e$. The typical 
behavior of them is illustrated in Fig.~\ref{rtcof}. We show the effect 
of the electric charge $q_e$ on the transmission and the reflection 
coefficients in case of both the scalar and electromagnetic field 
perturbations (cf. Fig.~\ref{rtcof}). We find that an increase in the 
magnitude of electric charge $q_e$ decreases the transmission coefficient 
in both cases of the perturbations. On the other hand, we observe the 
opposite effect of electric charge $q_e$ on the reflection coefficient 
when compared to the transmission coefficient. It turns out that the 
reflection coefficient increases with increasing magnitude of the 
electric charge $q_e$.

\section{Connection between shadow radius and QNMs}
\label{QNMs-shad}
Here we investigate a relationship between the black hole shadow and 
the real part of QNMs frequencies. To show this correspondence, let 
us first study the shadow of the 5D electrically charged Bardeen 
black holes using the geodesic approach. As we know from the symmetry 
of the spacetime \eqref{metric}, it admits three Killing vectors, 
namely $\partial_t$, $\partial_\phi$, and $\partial_\psi$. The 
presence of these Killing vectors can give rise to the associated 
conserved quantities, namely, the energy $E$, and the two angular 
momenta $L_{\phi}$ and $L_{\psi}$ in $\phi$ and $\psi$ directions, 
respectively. By using these conserved quantities and their relations 
with conjugate momenta, we can easily have the geodesic equations
\begin{eqnarray}
	\frac{d t}{d\sigma} &=& \frac{E}{f(r)}, \notag\\ [2mm]
	\frac{d \phi}{d\sigma} &=& \frac{L_{\phi}}{r^2 \sin^2 \theta}, 
	\notag\\ [2mm]
	\frac{d \psi}{d\sigma} &=& \frac{L_{\psi}}{r^2 \cos^2 \theta},
	\label{eom}
\end{eqnarray}
where $\sigma$ is an affine parameter. Apart from these geodesic 
equations, the radial and the angular geodesic equations can be 
derived by using the Hamilton-Jacobi equation 
\begin{eqnarray}
	\frac{\partial S}{\partial \sigma} = - \frac{1}{2}g^{\mu\nu} 
	\frac{\partial S}{\partial x^\mu} \frac{\partial S}{\partial x^\nu},
	\label{hmj}
\end{eqnarray}
where $S$ being the Jacobi action. The Jacobi action $S$ can be 
separated in the following form
\begin{eqnarray}
	S = \frac{1}{2} m_0^2 \sigma - E t + L_{\phi} \phi + L_{\psi} \psi 
	+ S_{r}(r) + S_\theta (\theta), \label{jac}
\end{eqnarray}
where $S_r$ and $S_\theta$ are functions of $r$ and $\theta$ only, 
respectively. The parameter $m_0$ denotes the mass of the test particle 
that vanishes in the photon's case. We now substitute \eqref{jac} into 
\eqref{hmj}, and after some straightforward calculations, we obtain
\begin{eqnarray}
	r^2 \frac{d r}{d\sigma} &=& \pm \sqrt{\mathcal{R}(r)}, 
	\notag \\ [2mm]
	r^2 \frac{d \theta}{d\sigma} &=& \pm \sqrt{\Theta(\theta)}, 
	\label{r-th-eom} 
\end{eqnarray}
where the function $R(r)$ and $\Theta(\theta)$ reads simply
\begin{eqnarray}
	\mathcal{R}(r) &=& E^2 r^4 -r^2 f(r) 
	\left(\mathcal{K} + L_{\phi}^2 + L_{\psi}^2\right), 
	\notag \\ [2mm]
	\Theta(\theta) &=& \mathcal{K} - L_{\phi}^2 \cot^2 \theta 
	- L_{\psi}^2 \tan^2 \theta. \label{R-Th}
\end{eqnarray}
Here $\mathcal{K}$ denotes the Carter constant which appears when we 
separate the coefficients of $r$ and $\theta$ during the Hamilton-Jacobi 
formulation.

In order to describe the black hole shadow, we introduce the 
celestial coordinates which in case of the 5D black holes are given 
\cite{Amir:2017slq} as following:
\begin{eqnarray}
	x &=& -\lim_{r_0 \rightarrow \infty} r_0
	\frac{p^{\hat{\phi}}+p^{\hat{\psi}}}{p^{\hat{t}}}, \notag \\ [2mm]
	y &=& \lim_{r_0 \rightarrow \infty}	r_0
	\frac{p^{\hat{\theta}}}{p^{\hat{t}}}, \label{celest}
\end{eqnarray}
where $p^{\hat{i}}$ being the contravariant components of the momenta in 
new coordinate basis. These contravariant components of the momenta can be 
easily computed by using the orthonormal basis vectors for the local 
observer \cite{Johannsen:2015qca,Amir:2017slq}, they become
\begin{eqnarray}
	p^{\hat{t}} &=& \frac{E}{f(r)}, \quad 
	p^{\hat{\phi}} = \frac{L_{\phi}}{r \sin \theta}, \quad
	p^{\hat{\psi}} = \frac{L_{\psi}}{r \cos \theta}, \notag \\ [2mm]
	p^{\hat{r}} &=& \pm \sqrt{f(r) \mathcal{R}(r)},\quad
	p^{\hat{\theta}} = \pm \frac{\sqrt{\Theta(\theta)}}{r}. \label{thrmom}
\end{eqnarray}
Plunging \eqref{thrmom} into \eqref{celest} and taking the limit 
${r_0 \rightarrow \infty}$, provides
\begin{eqnarray}
	x &=& -\left( \xi_{\phi} \csc \theta +\xi_{\psi} \sec \theta \right), 
	\notag \\ [2mm]
	y &=& \pm \sqrt{\eta -\xi_{\phi}^2 \cot^2 \theta 
	-\xi_{\psi}^2 \tan^2 \theta}, 
	\label{celestcord}
\end{eqnarray}
where we introduce the new quantities, $\xi_{\phi} =L_{\phi}/E$, 
$\xi_{\psi} =L_{\psi}/E$, and $\eta =\mathcal{K}/E^2$. These quantities 
are also known as the impact parameters. We now derive the limiting cases 
of \eqref{celestcord} according to the location of the observer. 
If the observer is situated in 
equatorial plane ($\theta_0 = \pi/2$), then the angular momentum 
$L_{\psi} = 0$, which implies $\xi_{\psi} = 0$, thus we have
\begin{eqnarray}
	x = -\xi_{\phi} , \quad	y = \pm \sqrt{\eta }.
	\label{celestcord1}
\end{eqnarray}
On the other hand, when $\theta_0 = 0$, in this case $L_{\phi} = 0$, 
which implies $\xi_{\phi} = 0$, thereby
\begin{eqnarray}
	x = -\xi_{\psi} , \quad	y = \pm \sqrt{\eta}.
	\label{celestcord2}
\end{eqnarray}
Equations \eqref{celestcord1} and \eqref{celestcord2} represent
a connection between the celestial coordinates and the impact parameters 
when the observer is situated at inclination angles $\theta_0=\pi/2$ and 
$\theta_0=0$, respectively. These equations are crucial to extract the 
information regarding the shadows of the black holes.

We further derive a relationship between the impact parameters of the 
spacetime \eqref{metric} by using the unstable spherical photon orbits 
conditions, $\mathcal{R}=0$ and $d\mathcal{R}/dr=0$, which turns out to be
\begin{eqnarray}\label{eqshad}
\eta + \xi_{\phi}^2 + \xi_{\psi}^2 = \frac{r^2}{f(r)}, \quad
	r = \frac{2 f(r)}{f^{\prime}(r)}, \label{impact}
\end{eqnarray} 
where prime ($\prime$) denotes the derivative with respect to $r$.
 \begin{figure}
    \includegraphics[scale=0.46]{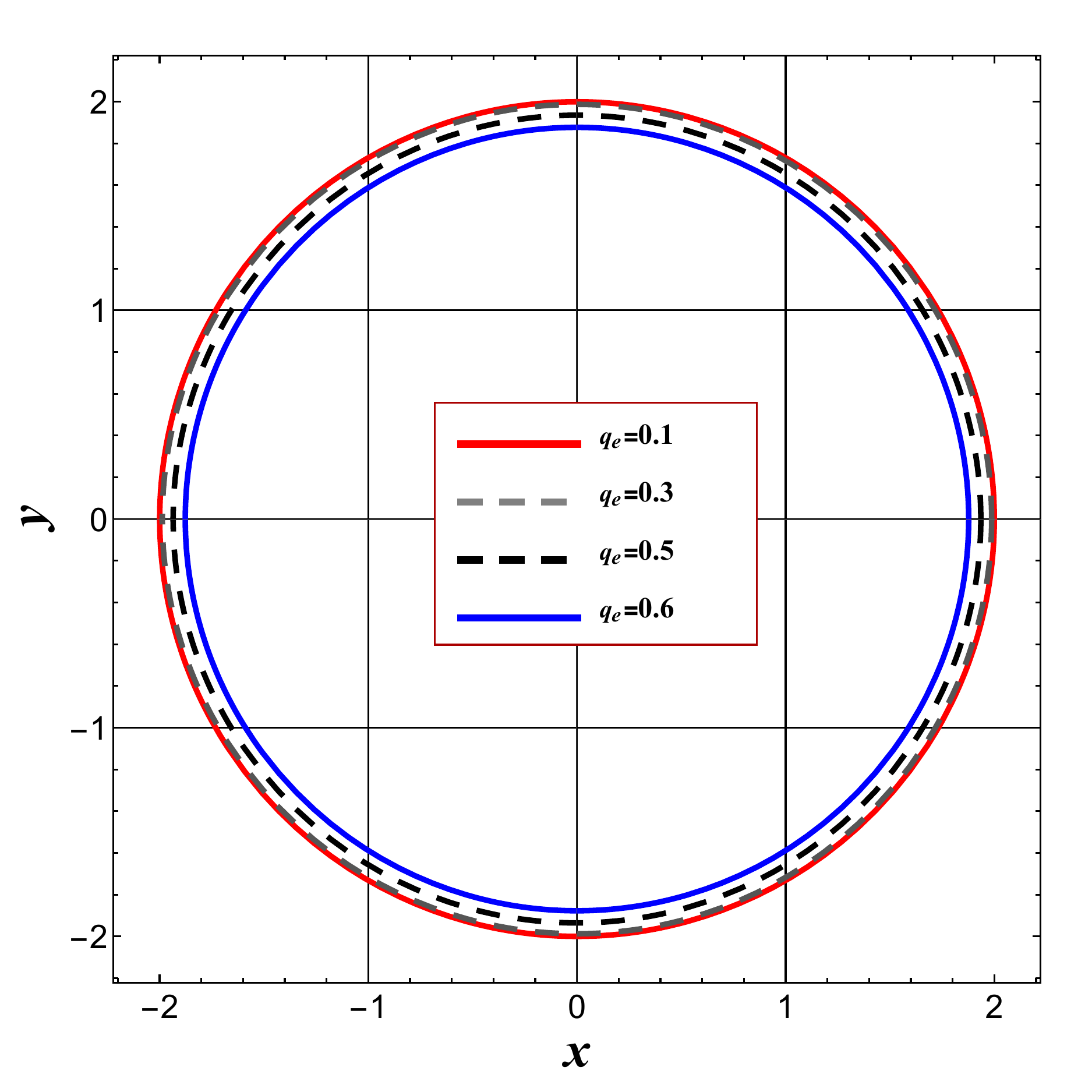}
    \caption{Illustration of shadows of the 5D electrically charged 
    Bardeen black holes for different values of electric charge $q_e$ 
    ($\mu=1$). \label{shad}}
 \end{figure}
Note that the metric function $f(r)$ is defined in \eqref{mf} and the 
derivative of it with respect to $r$ reads simply
\begin{eqnarray}
	f^{\prime}(r) = \frac{2 \mu (r^4-q_e^3 r)}
	{\left(r^3 +q_e^3\right)^{7/3}}.
\end{eqnarray}
On having all the related expressions with us, now we can portrait 
the shadows of the 5D electrically charged Bardeen black holes. The 
behavior of the shadows with electric charge $q_e$ can be seen in 
Fig.~\ref{shad}. We notice that the shape of the black hole shadow 
is a perfect circle. We see a decrease in radius of the black hole 
shadow with increasing magnitude of the electric charge $q_e$. 
Thus, we can conclude that the electric charge $q_e$ decreases 
the radius of the black hole shadow.

Now we proceed further to investigate the correspondence between the 
radius of the black hole shadow and the real part of the QNMs frequency. 
It has been argued in a seminal paper by Cardoso {\it et al.} 
\cite{Cardoso:2008bp} that in the eikonal limit, the real part of the 
the QNMs frequencies is related to the angular velocity of the unstable 
null geodesic. However, the imaginary part of the QNMs frequencies is 
associated to the Lyapunov exponent that determines the instability time 
scale of the orbits. This can be easily understood by the following 
equation \cite{Cardoso:2008bp} 
\begin{equation}
	\omega_{QNM}=\Omega_c l -i \left(n+\frac{1}{2}\right)|\lambda|, 
	\label{41}
\end{equation}
where $\Omega_c$ is the angular velocity at the unstable null 
geodesic, and $\lambda$ denotes the Lyapunov exponent. Furthermore, 
this correspondence is expected to be valid not only for the static 
spacetimes but also for the stationary ones. On the other hand, 
Stefanov {\it et al.} \cite{Stefanov:2010xz} showed a connection 
between the QNMs frequencies and the strong gravitational lensing of 
the spherically symmetric black holes spacetime. Most recently, one 
of the authors of this paper pointed out that the following relation 
relates the real part of the QNMs frequencies and the shadow radius 
(see for details \cite{Jusufi:2019ltj,Liu:2020ola})
\begin{equation}
	\omega_{\Re} = \lim_{l \gg 1} \frac{l}{R_s}, \label{k1}
\end{equation}
which is precise only in the eikonal limit having large values of 
multipole number $l$. Here $R_s$ denotes the radius of the black 
hole shadow. Hence, we can quickly rewrite the expression \eqref{41} 
as follows
\begin{equation}
	\omega_{QNM}=\lim_{l \gg 1} \frac{l}{R_s} 
	-i \left(n+\frac{1}{2}\right)|\lambda|.
\end{equation}
The importance of this correspondence relies on the fact that the 
shadow radius represents an observable quantity which can be measured 
by using direct astronomical measurement. Therefore, it is more 
convenient to express the real part of the QNMs frequencies in terms 
of the black hole shadow radius instead of the angular velocity. Another 
advantage of using \eqref{k1} is the possibility to determine the shadow 
radius once we have calculated the real part of QNMs and this, in turn, 
does not necessitate the use of the standard geodesic method. This 
close connection could be understood from the fact that the 
gravitational waves can be treated as massless particles propagating 
along the last null unstable orbit and out to infinity. It is thus 
expected that, in the eikonal limit, this correspondence could be valid 
for the scalar, the electromagnetic, and the gravitational field 
perturbations because they have the same behavior. Surprisingly, 
the link between the null geodesics and the QNMs frequencies is 
shown to be violated in the Einstein-Lovelock theory, even in the 
eikonal limit \cite{Konoplya:2017wot}. We therefore say that the 
correspondence between the real part of the QNMs frequencies and 
the shadow radius is not guaranteed for the gravitational field 
perturbations in the Einstein-Lovelock theory. Although the 
relation~\eqref{k1} is accurate only for large $l$, this relation 
can provide valuable information regarding the effect of the electric 
charge $q_e$ on the shadow radius even for small $l$. We can illustrate 
this fact using the numerical results of the real part of the QNMs 
frequencies presented in Tables~\ref{table1}, \ref{table2} and 
\ref{table3}, respectively. As we have already seen that the real 
part of the QNMs frequencies $\omega_{\Re }$ increases with the 
electric charge $q_e$. Therefore, if we use the inverse relationship 
between $\omega_{\Re }$ and the shadow radius $R_S$,
\begin{equation}
	\omega_{\Re }(q_e) \propto \frac{1}{R_s(q_e)}, 
\end{equation}
\begin{figure}
   	\includegraphics[width=7.8 cm]{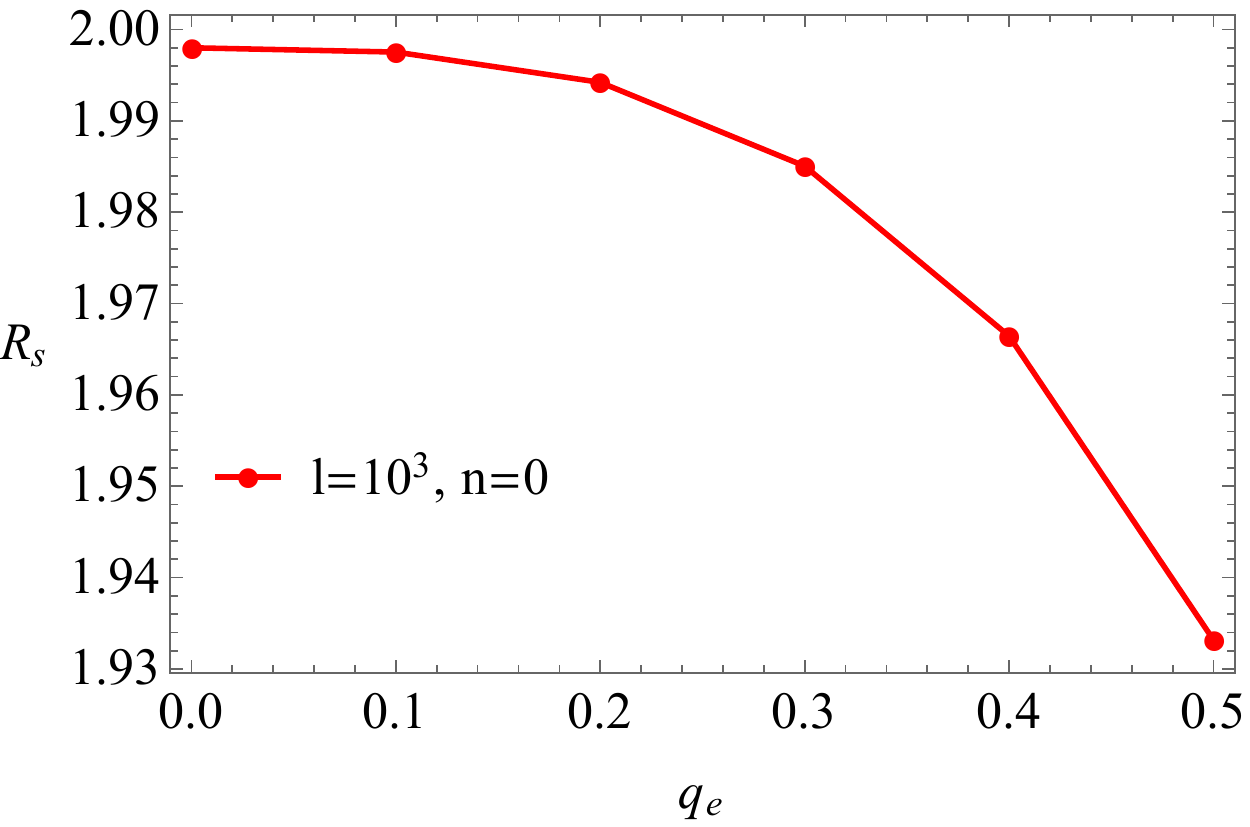}
    \caption{Shadow radius of the 5D electrically charged Bardeen 
    black holes as a function of electric charge $q_e$ obtained via 
    the relation \eqref{k1} with mass $\mu=1$.  \label{rs}}
\end{figure}
it follows that the shadow radius should decrease by increasing the 
magnitude of the electric charge $q_e$. This effect indeed is shown 
to be the case as depicted in Fig.~\ref{shad} by using the geodesic 
method. It is quite surprising that we can deduce this information 
directly from the inverse relationship between the real part of the 
QNMs frequencies and the shadow radius \eqref{k1}, even in case of 
the small multipole number $l$. However, the relation \eqref{k1} 
is precise only in the eikonal regime having $l>>1$. We use relation 
\eqref{k1} to demonstrate the effect of electric charge $q_e$ on the 
shadow radius (cf. Fig.~\ref{rs}). We also discover that the shadow 
radius of the 5D electrically charged Bardeen black holes is smaller 
than that of the 5D Schwarzschild-Tangherlini black holes. 
On the other hand, sometimes we use the inverse setup; 
namely, we can evaluate numerically the shadow radius $R_s$ from the 
geodesic equations to find the real part of QNMs frequencies. To 
illustrate this fact, we use the correspondence between the shadow 
radius of black hole and the real part of QNMs frequencies to the 
sub-leading regime to half of its value for the $D$-dimensional case 
reported recently in  \cite{Cuadros-Melgar:2020kqn}:
\begin{equation}\label{modeq}
    \omega_{\Re}=R_s^{-1}\left(l+\frac{D-3}{2}\right).
\end{equation}
It is clear from \eqref{modeq} that in large angular momentum 
regime, i.e., $l>>1$, we can recover \eqref{k1}. Again, this 
correspondence is usually accurate in the eikonal regime, but in 
some cases, it is also accurate even for small multipole number $l$. 
To verify this, we use the shadow radius expressed in (\ref{eqshad}) 
in order to compute the real part of QNMs frequencies from 
\eqref{modeq}. In Table \ref{table4}, we present numerical results 
of the shadow radius and the real part of QNMs frequencies for a 
given multipole number $l$ by varying the magnitude of electric 
charge $q_e$. When we compare the numerical results presented in Table 
\ref{table4} with the results obtained using the WKB method for the 
scalar (Table~\ref{table1}) and electromagnetic field perturbations 
(\ref{table2} and \ref{table3}), we see that for small $l$, the 
accuracy and precision between two methods is higher for the scalar 
field perturbations. In the case of electromagnetic field perturbations, 
we observe a higher accuracy for modes \textbf{I} in comparison to 
modes \textbf{II}. Indeed, the accuracy and precision of the above 
two methods increase with the multipole number $l$.
\begin{table}[tbp]
    \caption{ \label{table4} Numerical values of the shadow radius and the 
    real part of QNMs frequencies obtained via \eqref{modeq}.}
        \begin{tabular}{|l|l|l|l|l|l|}
        \hline
    \multicolumn{1}{|c|}{ } &  \multicolumn{1}{c|}{  $l=1, n=0$ } 
    & \multicolumn{1}{c|}{  $l=2, n=0$ } &   \multicolumn{1}{c|}{  $l=3, n=0$ } 
    & \multicolumn{1}{c|}{  $l=4, n=0$ } & \multicolumn{1}{c|}{} \\
    \hline
  $q_e$ & \quad $\omega_{\Re}$ & \quad $\omega_{\Re}$ & \quad $\omega_{\Re}$  
  & \quad $\omega_{\Re}$   & \quad $R_s$  \\ 
        \hline
0 & 1.0 & 1.5  & 2.0  &  2.5    &  2.0 \\ 
0.1 & 1.000235918 & 1.500353876 & 2.000471835  & 2.500589794  & 1.999528276 \\ 
0.2 & 1.001899545 & 1.502849318 & 2.003799091  & 2.504748864  & 1.996208112 \\
0.3 & 1.006526831 & 1.509790247 & 2.013053662  & 2.516317078  & 1.987030984 \\
0.4 & 1.016051116 & 1.524076674 & 2.032102232  & 2.540127790  & 1.968404905\\
0.5 & 1.033537114 & 1.550305671 &  2.067074228 & 2.583842785  & 1.935102255 \\
        \hline
        \end{tabular}
\end{table}

\section{Absorption cross-section}
\label{abcrosec}
\begin{figure}
	\includegraphics[width=8 cm]{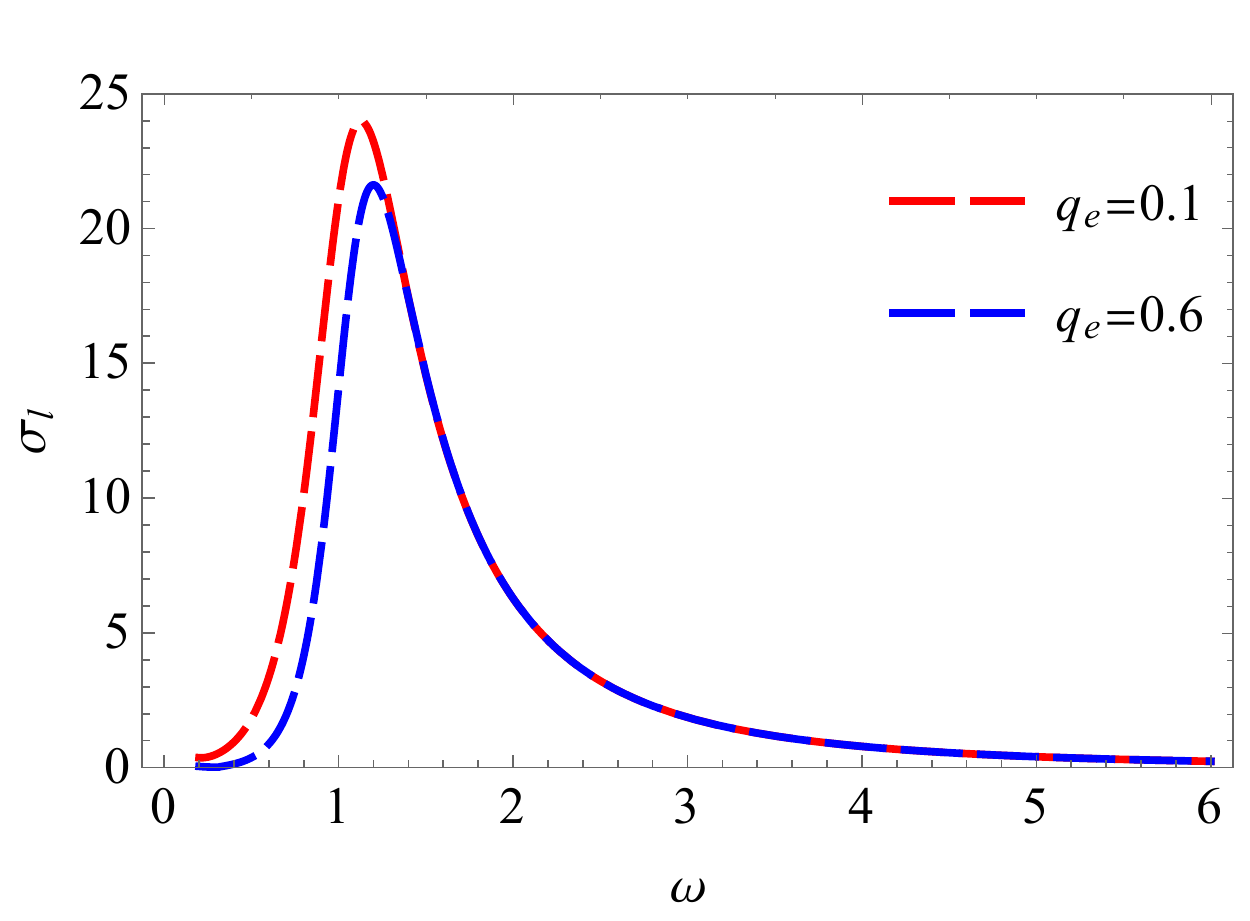}
	\caption{\label{scfcs} The plot shows the dependence of partial 
	absorption cross-section versus frequency of the scalar field 
	perturbations ($\mu=1$ and $l=1$). }
\end{figure}

\begin{figure*}
	\begin{center}
	\includegraphics[width=8 cm]{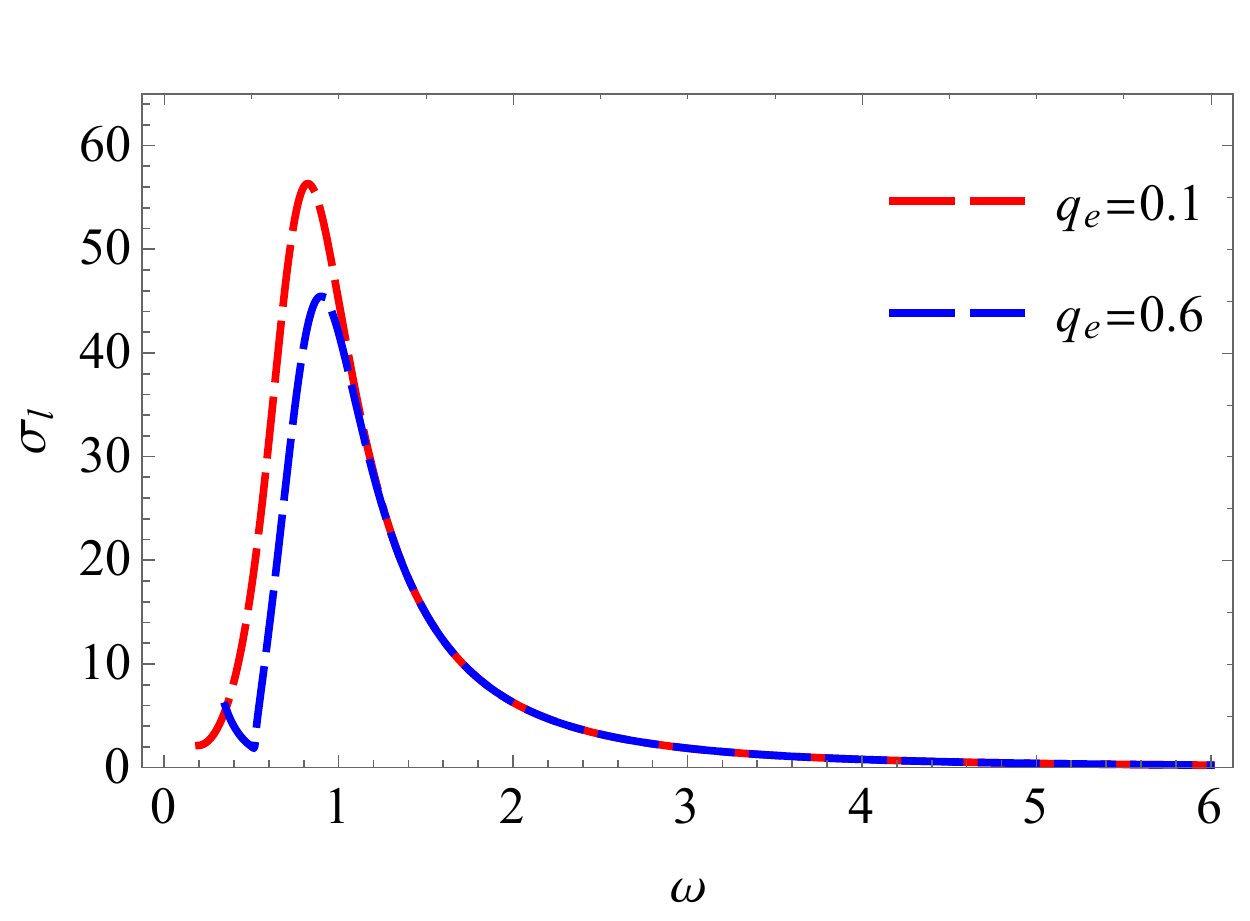}
	\includegraphics[width=8 cm]{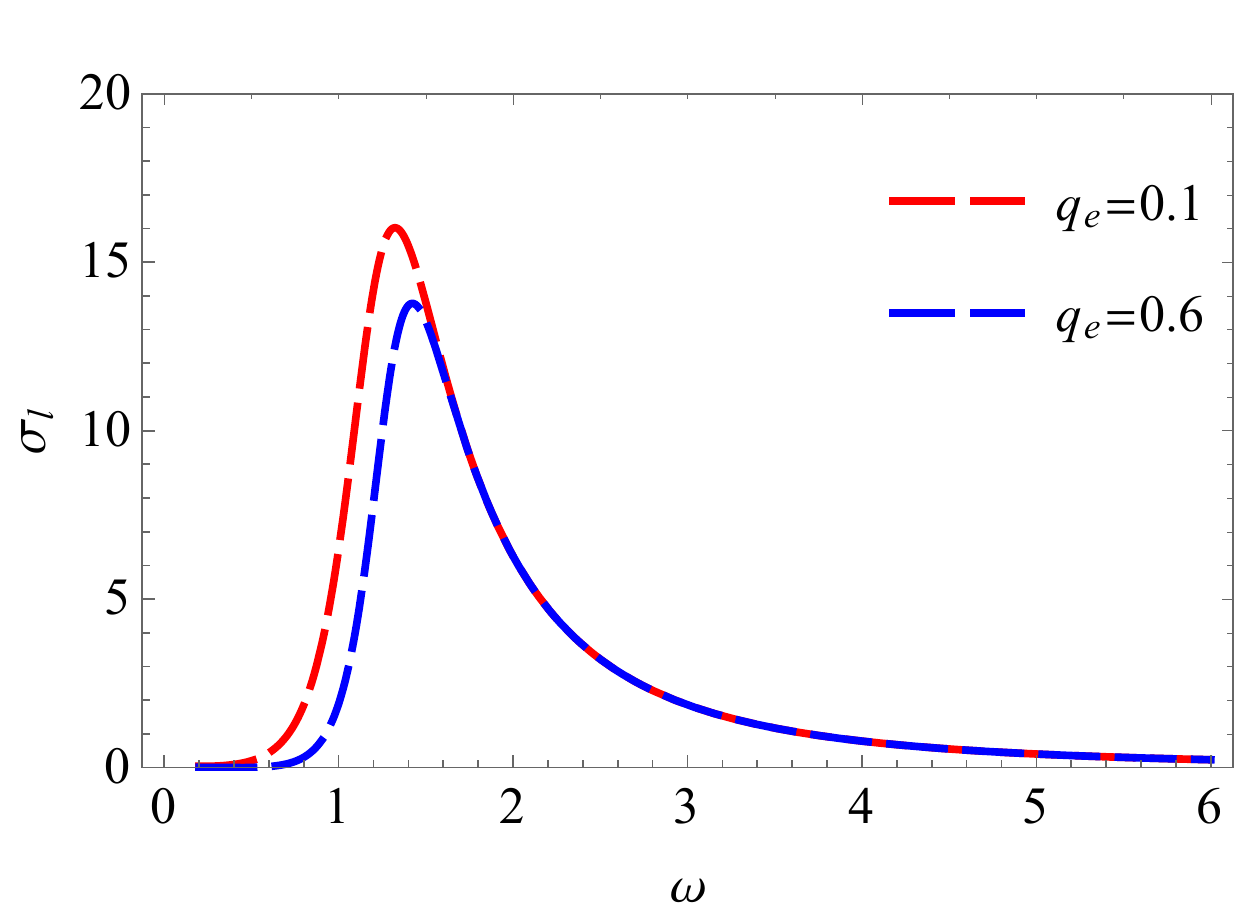}
	\caption{\label{emfcs} (Left panel) Plot shows the dependence of 
	partial absorption cross-section versus frequency in case of 
	modes \textbf{I} of the electromagnetic field perturbations. 
	(Right panel) Plot shows the dependence of partial absorption 
	cross-section versus frequency in case of modes \textbf{II} of 
	the electromagnetic field perturbations ($\mu=1$ and $l=1$).}
	\end{center}
\end{figure*}

In this section, we are going to determine the partial absorption 
cross-section in 5D electrically charged Bardeen black holes spacetime. 
This is another important quantity which in D-dimensions can be 
defined as following \cite{Decanini:2011xi}:
\begin{eqnarray}
    \sigma_l =  \frac{\pi^\frac{D-2}{2}}{\Gamma(\frac{D-2}{2})\, 
    \omega^{D-2}} \frac{(l+D-4)! \,(2l+D-3)}{l!} 
    \left|T_l(\omega)\right|^2.  
\end{eqnarray}
Thus, in case of the 5D electrically charged Bardeen spacetime, the 
partial absorption cross-section reads simply
\begin{eqnarray}
    \sigma_l = \frac{4 \pi (l+1)^2}{\omega^3} \left|T_l(\omega)\right|^2.
\end{eqnarray}
Furthermore, it could be useful to define the total absorption 
cross-section, which can be given by
\begin{eqnarray}
     \sigma =\frac{\pi^\frac{D-2}{2}}{\Gamma(\frac{D-2}{2})\,
     \omega^{D-2}} \sum_{l=0}^{\infty} \frac{(l+D-4)! \,(2l+D-3)}{l!} 
     \left|T_l(\omega)\right|^2.
\end{eqnarray}
In Figs.~\ref{scfcs} and \ref{emfcs}, we plot the partial absorption 
cross-section versus $\omega$ for the scalar and electromagnetic 
field perturbations, respectively. From these plots, we can see that 
the partial absorption cross-section decreases by increasing the 
magnitude of the electric charge $q_e$. In particular, for a given 
value of the electric charge $q_e$, we see that the partial absorption 
cross-section initially increases with an increasing value of $\omega$. 
It reaches the maximum value at some critical value of $\omega$. 
Then it decreases in the limit of large $\omega$ since the partial 
absorption cross-section behaves like the inverse of $\omega^{3}$.

In high energy scale regime, the wavelength is considered to be 
negligible relative to the horizon scale of the black hole. 
Hence, one could identify the classical capture cross-section of 
the light rays with the geometric cross-section of the light rays 
given by the the expression \cite{Decanini:2011xi,Toshmatov:2016bsb}
\begin{eqnarray}
    \sigma_{geom}=\frac{\pi^{\frac{D-2}{2}} b_{ps}^{D-2}}{\Gamma(D/2)},
\end{eqnarray}
where $b_{ps}$ is the critical impact parameter of the light rays. 
It can be defined as the ratio of the angular momentum and the 
photon's energy moving along the spherical photon orbits, i.e., 
$b_{ps}=J/E$. Now we can use the geometric-optics correspondence 
between the parameters of the QNMs frequencies and the conserved 
quantities along geodesics. In particular, the particle energy can 
be identified as the real part of the QNMs frequencies. Besides, 
the azimuthal quantum number corresponds to the angular momentum 
\cite{Yang:2012he}, hence, we can quickly express
\begin{equation}
    E \to \omega_{\Re},\quad \text{and} \quad J \to l.
\end{equation}
It means that we can identify the impact parameter of the light rays 
with the shadow radius, i.e., $b_{ps} \to R_s$. Further, we can use 
\eqref{k1} to express the geometric cross-section in terms of the real 
part of the QNMs frequency as follows
\begin{eqnarray}
    \sigma_{geom}=\lim_{l>>1}\frac{\pi^{\frac{D-2}{2}} l^{D-2}}{\Gamma(D/2)\omega_{\Re}^{D-2}}, \label{geomcs}
\end{eqnarray}
which is valid in the eikonal regime. As we have already pointed out 
that due to the symmetry of the scattering properties, the greybody 
factors are also present in the emission spectrum of the black holes. 
In other words, if the black hole emits in the eikonal regime, the 
same connection between the real part of the QNMs frequency and the 
geometric cross-section given by \eqref{geomcs} is to be expected. 
Basically, this means that one can calculate the geometric 
cross-section by using the real part of the QNMs frequencies. 
To the best of our knowledge, a relation between the geometric 
cross-section and the real part of the QNMs frequencies has not 
been explored before. Before we proceed to calculate the total 
absorption cross-section, let us point that in 
Ref. \cite{Decanini:2011xi}, it has been shown that there are 
fluctuations (of regular oscillations) of the high-energy 
(frequency) absorption cross-section around the limiting value 
of the geometric cross-section. In particular, it was found that 
the oscillatory part of the absorption cross-section of the 
massless scalar waves is given \cite{Decanini:2011xi} by
\begin{eqnarray}
    \sigma_{osc} = (-1)^{D-3}4 (D-2) \pi \lambda R_s e^{-\pi \lambda R_s}\,
    \sigma_{geom}\, \text{sinc}(2 \pi R_s \omega),
\end{eqnarray}
where we have used the correspondence, $b_{ps} \to R_s$. Moreover, the 
function has been introduced $\text{sinc}(x) = \sin(x)/x$ and $\lambda$ 
is the Lyapunov exponent used for the analysis of the instability of 
the null geodesics. This means that the total absorption cross-section 
of the massless scalar waves is the sum of the geometric and the 
oscillatory cross-sections
\begin{eqnarray}
    \sigma \approx \sigma_{geom}+\sigma_{osc}.
\end{eqnarray}
\begin{figure*}
	\begin{center}
	\includegraphics[width=8.5 cm]{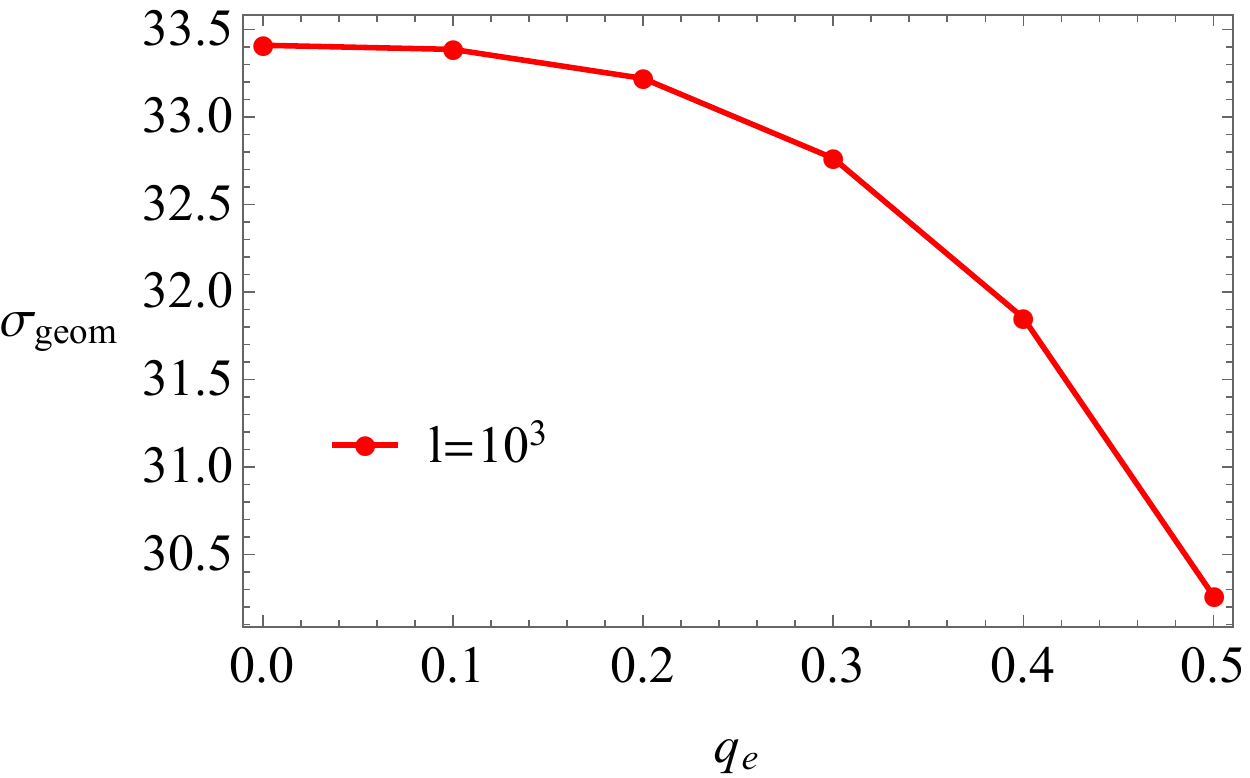}
	\includegraphics[width=7.8 cm]{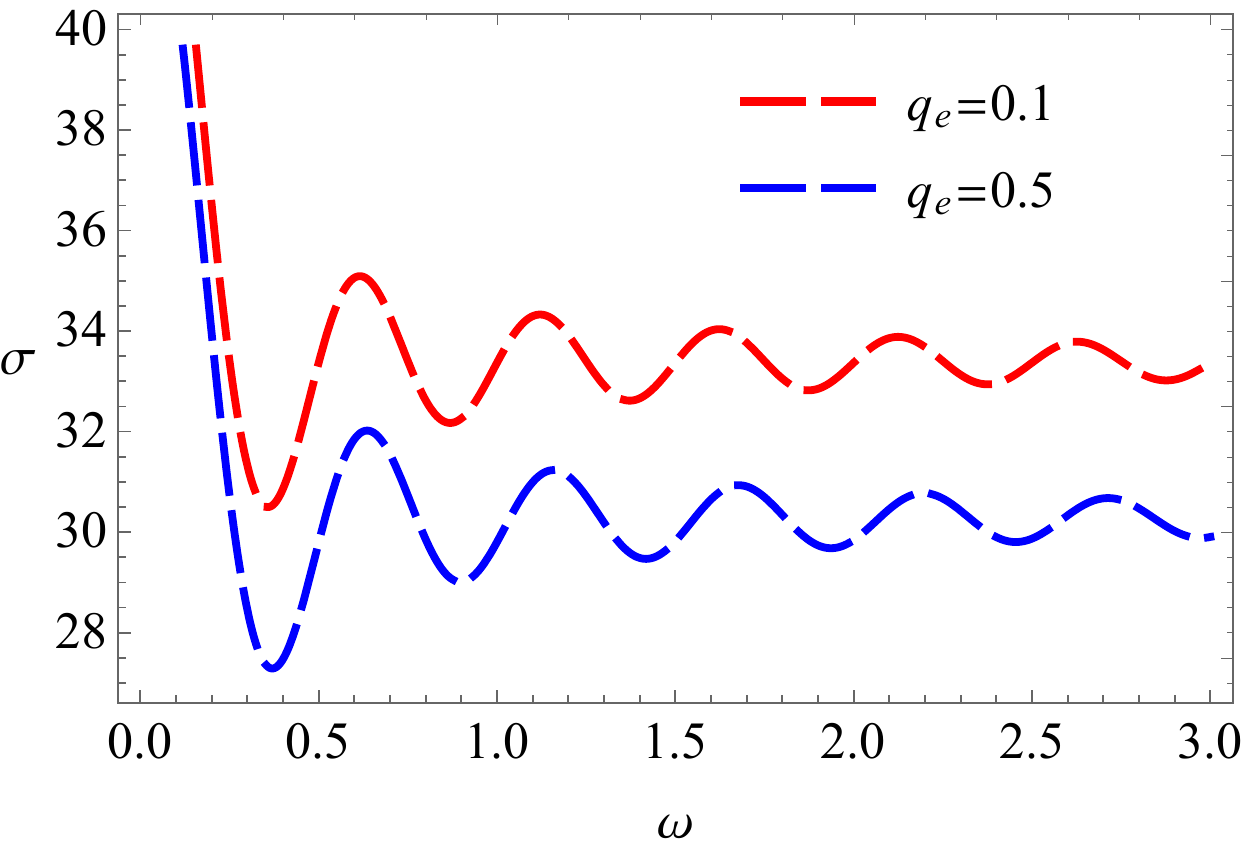}
	\caption{\label{cross} (Left panel) Geometric cross-section as a 
	function of electric charge $q_e$. In both plots we have used the 
	scalar field perturbation to calculate $\omega_{\Re}$ with $\mu=1$. 
	One can reach the same conclusion using the electromagnetic field 
	perturbations, provided $l>>1$. (Right panel) Plot of the total 
	absorption cross-section versus frequency of the scalar field 
	perturbations. We have used $q_e=0.1$ (red curve) and $q_e=0.5$ 
	(blue curve), respectively ($l=10^3$, $\lambda=0.8$ and $\mu=1$).}
	\end{center}
\end{figure*}
From Eq. \eqref{geomcs}, we can easily find that in case of the 5D 
black holes, the geometric cross-section can be expressed in terms 
of the real part of the QNMs frequency as follows
\begin{equation}
    \sigma_{geom}=\lim_{l \gg 1}  \frac{4 \pi l^3}{3\,\omega_{\Re}^3}. 
    \label{5Dgcs}
\end{equation}
In Fig.~\ref{cross} (left panel), we portrait the geometric 
cross-section as a function of the electric charge $q_e$. We see that 
the geometric cross-section decreases with an increase in the magnitude 
of the electric charge $q_e$. We find that in the eikonal limit, the 
geometric cross-section and the shadow radius share a similar behavior 
with respect to the electric charge $q_e$ which can be explained from 
the fact that the geometric cross-section scales with the shadow radius. 
On the other hand, the oscillatory part in case of the 5D black holes 
reads simply
\begin{eqnarray}
    \sigma_{osc}=12 \pi \lambda R_s e^{-\pi \lambda R_s}\,
    \sigma_{geom}\, \text{sinc}(2 \pi R_s \,\omega).
\end{eqnarray}
In Fig.~\ref{cross} (right panel), we show the total absorption 
cross-section as a function of $\omega$ and fixed values of the 
electric charge $q_e$ in case of the scalar field perturbations. 
We see that the total absorption cross-section decreases by 
increasing the electric charge $q_e$. In this sense, our result 
is consistent with effect of the electric charge in case of the 
4D electrically charged black hole and generalizes the results 
reported in \cite{Decanini:2011xi}.

\section{Conclusion}
\label{concl}
In this work, we have performed a comprehensive discussion on QNMs in 5D 
electrically charged Bardeen black holes spacetime. The study has provided 
impressive results in five dimensions when considering general relativity 
coupled to the nonlinear electrodynamics. To compute the QNMs frequencies, 
we have discussed the scalar and the electromagnetic field perturbations. 
We have used the WKB approach up to sixth order to determine the 
numerical values of the QNMs frequencies. The effect of nonlinear 
electric charge $q_e$ on the QNMs frequencies has been investigated,
which provides a significant impact. In studying the scalar field 
perturbations, it has been found that the electric charge $q_e$ increases 
the real part of the QNMs frequencies and decreases the imaginary part of 
the QNMs frequencies in an absolute amount. While discussing the 
electromagnetic field perturbations, we discovered that an increase 
in the electric charge $q_e$ increases the real part of the QNMs 
frequencies and decreases the imaginary part of the QNMs frequencies 
in absolute value in both cases of the modes \textbf{I} and 
\textbf{II}, respectively. We have noticed that the numerical values 
of the QNMs frequencies are higher in the scalar field perturbations 
compared to the electromagnetic field perturbations. This indicates that 
the scalar field perturbations oscillate more rapidly in comparison to 
the electromagnetic field perturbations. On the other hand, the scalar 
field perturbations damp more quickly than the electromagnetic field 
perturbations.

In further analysis, we have discussed the scattering and the greybody 
factors. We calculated analytical expressions of the reflection and 
the transmission coefficients. It has been found that the reflection 
coefficient increases with electric charge $q_e$ while the transmission 
coefficient decreases with electric charge $q_e$. We extended our 
analysis to determine a direct connection between the black holes' 
shadows and the QNMs frequencies. The null geodesics and spherical 
photon orbits conditions have been discussed to describe the shadow 
of black holes. The black holes' shadows have been portrayed by 
varying electric charge $q_e$, and its typical behavior, has also been 
discussed. We have found that the presence of electric charge $q_e$ affects 
the radius of the black hole shadow, which decreases by an increase in the 
magnitude of electric charge $q_e$. We have shown that this result can be 
obtained by means of the real part of the QNMs frequencies valid in the 
eikonal limit. We have also discussed how the electric charge $q_e$ affects 
the partial absorption cross-section in case of 5D electrically charged 
black holes. Finally, we have explored the total absorption cross-section 
in the high energy scale, which consists of the geometric cross-section 
and oscillatory parts. Importantly, we have expressed the geometric 
cross-section in terms of the real part of the QNMs frequencies by using 
the geometric-optics correspondence. Our result shows that the total 
absorption cross-section decreases with an increase in the magnitude of 
electric charge $q_e$.

Our work can be extended to AdS spaces as they open a new avenue 
on the onset of the connection of the photon orbits to the thermodynamic 
phase transition of the AdS black holes. On the other hand, the black hole 
shadows and the QNMs frequencies of the AdS black holes are connected with 
the stability, the thermodynamic phase transition, and the hydrodynamic 
region of strongly coupled field theories. Hence, the study of the QNMs 
in the 5D electrically charged regular AdS black holes and its connection 
to the shadow properties will be a natural extension of our work in a 
future project.

\acknowledgements
MA acknowledges that this research work is supported by the National Research 
Foundation, South Africa. MSA's research is supported by the ISIRD grant 
9-252/2016/IITRPR/708. SDM acknowledges that this work is based on research 
supported by the South African Research Chair Initiative of the Department 
of Science and Technology and the National Research Foundation. We are 
indebted to the anonymous referee for useful comments and suggestions.

\bibliographystyle{apsrev4-1}
\bibliography{References}
   
\end{document}